\documentclass[twocolumn]{aastex631}
\usepackage{pifont}
\usepackage{mathtools}

\usepackage{graphicx}
\usepackage{amsmath}
\usepackage{natbib}
\usepackage{amssymb}

\begin{document}

%\title{Physical Properties of Synchrotron-Emitting Shocks at Arbitrary Velocities}
%\title{Relativistically-Correct Inferences on Synchrotron-Emitting Shocks}
%\title{Inferring the Physical Properties of Mildly-Relativistic Synchrotron-Emitting Shocks}
\title{The Peak Frequency and Luminosity of Synchrotron Emitting Shocks:

from Non-Relativistic to Ultra-Relativistic Explosions
}

\shorttitle{Peak Properties of Synchrotron Shocks}
\shortauthors{Margalit \& Quataert}

\correspondingauthor{Ben Margalit}
\email{margalit@umn.edu}

\author[0000-0001-8405-2649]{Ben Margalit}
%\affiliation{Minnesota Institute for Astrophysics, University of Minnesota, Minneapolis, MN 55455, USA}
\affiliation{School of Physics and Astronomy, University of Minnesota, Minneapolis, MN 55455, USA}
%\affiliation{Astronomy Department and Theoretical Astrophysics Center, University of California, Berkeley, Berkeley, CA 94720, USA}

\author[0000-0001-9185-5044]{Eliot Quataert}
\affiliation{Department of Astrophysical Sciences, Princeton University, Princeton, NJ 08544, USA} 



\begin{abstract}
Synchrotron emission is ubiquitous in explosive astrophysical events---it is a natural byproduct of shocks formed when matter expelled by the explosion collides with ambient material. This emission is well-observed in various classes of transients, and is often interpreted within a canonical `equipartition' framework that allows physical properties of the shock to be inferred from the frequency and luminosity at which the observed spectral energy distribution (SED) peaks. This framework has been remarkably successful in explaining observations of radio supernovae. It has also been used for trans-relativistic explosions, where the shock velocities approach the speed of light. However, the conventional framework does not incorporate relativistic effects. Neither does it account for thermal electrons, which have been shown to be important for high-velocity shocks. In this paper we describe a revised framework that accounts for these two effects, and is applicable to non-relativistic, trans-relativistic, and ultra-relativistic explosions. We show that accounting for these effects can dramatically change the inferred parameters of high-velocity shocks, and in particular---that the shock velocity, ambient density, and total energy are overestimated by the conventional non-relativistic framework. We delineate the phase-space where such modifications are important in terms of observationally measurable parameters. We also find a novel upper limit on the peak synchrotron luminosity of shock-powered transients, which is remarkably consistent with existing observations. Finally, we discuss a prediction of the model---that the SED will qualitatively change as a function of shock velocity---and show that this is broadly consistent with data for representative events (e.g., SN1998bw, AT2018cow, CSS161010, AT2020xnd).
\end{abstract}



\section{Introduction}

Transient astrophysical events produce emission that spans the entire electromagnetic spectrum and can be accompanied by multi-messenger counterparts in the form of neutrinos, gravitational-waves, and cosmic rays. These events are often the harbingers of extreme and energetic ``explosions'' such as core-collapse and thermonuclear supernovae (SNe), gamma-ray bursts (GRBs), compact object mergers, or tidal disruption events (TDEs). In these cases and in many other scenarios, the explosive event expels some material
%---typically called an ejecta---
at relatively high velocities. 
The ejected material will inevitably collide with the ambient medium in the vicinity of the source and form a shock. Such shocks are well known sources of broadband synchrotron emission that can be detected at different frequencies (most commonly, in the radio).

Well-studied examples include the classes of radio SNe and GRB afterglows \cite[e.g.,][]{Weiler+02}.
In both cases, it is usually assumed that observed emission is produced by a population of non-thermal relativistic electrons accelerated at the shock front through a first-order Fermi process. These electrons produce synchrotron emission by gyrating in a magnetic field that is typically assumed to be amplified self-consistently by the (collisionless) shock. This model has worked remarkably well in explaining observations of ``normal'' radio SNe and GRB afterglows \citep[e.g.,][]{Chevalier82,Weiler+86,Chevalier98,Sari+98}. It is also broadly consistent with expectations from theory and kinetic simulations \citep[e.g.,][]{Bell78,BlandfordEichler87,Blandford&Ostriker78,Spitkovsky08,Sironi&Spitkovsky09,Sironi&Spitkovsky11,Sironi+13,Park+15}.

In recent years, interest in synchrotron-emitting shock-powered transients has been renewed with the discovery of a growing sample of mildly- and trans-relativistic explosions including broad-lined Ic SNe (Ic-BL) and the emerging class of Fast Blue Optical Transients (FBOTs) which are most known by their prototypical event AT2018cow \citep{Margutti+19,Ho+19}.
\cite{Margalit&Quataert21} showed that for such events, observed synchrotron emission could instead be powered by the population of electrons that are thermalized in the post-shock region (rather than non-thermal electrons, which must undergo an additional acceleration process). This thermal-electron model reproduced well the observational data for several FBOTs \citep{Ho+22}, and is a natural expectation for shocks whose velocity exceed $\beta_{\rm sh}c \gtrsim 0.2c$ \citep{Margalit&Quataert21}.
The contribution of thermal electrons to observed synchrotron emission has also been discussed in the GRB literature \citep{Giannios&Spitkovsky09,Warren+17,Ressler&Laskar17,Warren+18,Samuelsson+20,Warren+22}. 
Still, the importance of synchrotron emission from thermal electrons in shock-powered transients is not yet widely appreciated.

In this paper, we extend our previous work \citep{Margalit&Quataert21} by incorporating relativistic effects and developing an analytic formalism for shock-powered synchrotron emission (and absorption) that is correct for arbitrary shock velocities. This allows us to properly survey the full parameter-space from non-relativistic radio SNe through trans-relativistic FBOTs and SNe Ic-BL, to ultra-relativistic GRBs, all within a single framework.
A particular focus of this paper is applying this formalism to describe the frequency and luminosity at which the spectral energy distribution (SED) of the resulting synchrotron emission peaks. 

In a canonical paper, \cite{Chevalier98} showed that the radio data of typical radio SNe is consistent with a synchrotron self-absorption (SSA) model in which the SED is self-absorbed at low frequencies and the peak  of the SED coincides with the SSA frequency (that is, the frequency at which the SSA optical depth is $\sim 1$). Furthermore, \cite{Chevalier98} showed that---in cases where this model applies---observations of the frequency $\nu_{\rm pk}$ and specific luminosity $L_{\nu_{\rm pk}}$ at which the SED peaks are sufficient to fully determine the physical properties of the shock. In particular, \cite{Chevalier98} show how observed values of $\nu_{\rm pk}$, $L_{\nu_{\rm pk}}$, and the time $t$ at which they are observed (measured from the explosion epoch) can be used to infer the shock radius and post-shock magnetic field. These can subsequently be used to determine all other physical variables of the system, including the shock velocity, the density of the ambient medium, and the explosion energy.
The method described by \cite{Chevalier98} has been used extensively throughout the literature, and with great success. However, this formalism is inapplicable when the shock velocities are high. In particular, the \cite{Chevalier98} analysis does not include relativistic corrections which are important when the proper-velocity of the shock is $(\Gamma\beta)_{\rm sh} \gtrsim 1$. It also neglects the contribution of thermal electrons that has been shown to be important when $\beta_{\rm sh} \gtrsim 0.2$ \citep{Margalit&Quataert21}.
The aim of the present work is to extend this formalism such that it can be appropriately applied to shocks of arbitrary velocity.

Our primary result is a set of expressions that can be used to infer the shock velocity and ambient density from observed values of $\nu_{\rm pk}$, $L_{\nu_{\rm pk}}$, and $t$. These can subsequently be used to find the shock radius, post-shock magnetic field, and energy (see \S\ref{sec:implications} and Appendix~\ref{sec:Appendix}), and are valid for any shock velocity.
We begin by presenting our formalism in \S\ref{sec:model}. In \S\ref{sec:SEDs} we describe the resulting synchrotron SEDs. The SED will appear qualitatively different depending on the shock velocity, and is separated into three regimes: (a) the SED peak is set by SSA and non-thermal (power-law) electrons dominate emission; (b) SSA governs the SED at peak but thermal electrons dominate emission at this frequency; (c) emission at the peak of the SED is optically-thin. These regimes are treated in \S\ref{sec:powerlaw}, \S\ref{sec:thermal}, and \S\ref{sec:optically-thin}, respectively. We discuss implications of our results and compare to observations in \S\ref{sec:implications}. Finally, we summarize our findings and conclude in \S\ref{sec:summary}.

\section{The Physical Model}
\label{sec:model}

We follow the non-relativistic framework of \cite{Margalit&Quataert21}, extending this model to arbitrary shock velocity. 
We additionally adopt an effective line-of-sight (LOS) approximation in which flux is considered to be dominated by material at some representative point within the source.
For relativistically expanding sources, emission from different regions will reach the observer at different times. A complete calculation of the observed flux in this case therefore necessitates a numerical approach and requires knowledge of the shock expansion history. This has been performed numerically for the case of GRBs assuming Blandford-McKee expansion \citep{Granot+99a,Granot+99b}, or as post-processing of hydrodynamic simulations \citep[e.g.,][]{vanEerten+12}. Wishing to avoid imposing any specific expansion history (which may differ in different contexts and for different shock velocities), and in the interest of obtaining analytically-tractable results---we adopt the LOS approximation.
As part of this approximation we take the effective Doppler factor to be $\mathcal{D} = \Gamma$, where $\Gamma = 1/\sqrt{1-\beta^2}$ is the Lorentz factor of post-shock emitting matter with velocity $\beta c$. 
We also assume that the shock radius (as measured in the lab-frame) is
\begin{equation}
    \label{eq:R}
    R 
    = \sqrt{ 1 + \left(\Gamma\beta\right)_{\rm sh}^2 } \left(\Gamma\beta\right)_{\rm sh} c \bar{t}
    ~;~~~ \bar{t} \equiv \ell_{\rm dec} t
\end{equation}
where $t$ is the elapsed time measured in the lab (observer) frame
and $\ell_{\rm dec} \gtrsim 1$ is a dimensionless parameter that accounts for potential deceleration of the shock.
Here $\Gamma_{\rm sh}$ is the shock Lorentz factor and $\beta_{\rm sh} c$ its corresponding velocity,
and we have expressed $R$ in terms of the proper velocity $\left(\Gamma\beta\right)_{\rm sh} = \sqrt{\Gamma_{\rm sh}^2-1}$.
The shock velocity is related to the post-shock velocity via the Blandford-McKee relations (\citealt{BlandfordMcKee76}; their Equation~5)\footnote{ 
This relationship is exact for arbitrary velocities and does not depend on the shock's evolution history (in particular, it does not assume that the shock evolves according to the ultra-relativistic Blandford-McKee self-similar solution).
}
where we assume that the post-shock adiabatic index can be approximated by $\hat{\gamma} = (4+\Gamma^{-1})/3$ (see Equation~\ref{eq:Appendix_gb_of_gbsh}).

Equation~(\ref{eq:R}) reduces to $R = \beta_{\rm sh} c \bar{t}$ in the non-relativistic regime and to $R = 2 \Gamma^2 c \bar{t}$ in the ultra-relativistic limit. 
For $\ell_{\rm dec}=1$ it describes a constant velocity shock.
A decelerating shock with the same instantaneous velocity would have a larger radius,
and can be described by taking $\ell_{\rm dec} \gtrsim 1$.
For example, $\ell_{\rm dec} = 8$ for an ultra-relativistic shock decelerating according to the Blandford-McKee solution and viewed along the LOS. 
(\citealt{Sari97}; accounting for emission off the LOS leads to an effective $\ell_{\rm dec} \sim 3-7$; \citealt{Sari98})
For a Newtonian shock that expands as $R \propto t^m$ \citep[e.g.,][]{Chevalier98} we instead have $\ell_{\rm dec} = 1/m$. This implies
$1 \lesssim \ell_{\rm dec} \lesssim 3/2$ for the hydrodynamic solution described by  \cite{Chevalier82}
and $\ell_{\rm dec} = 5/2$ for the case of a Sedov-Taylor blast-wave propagating into a constant-density medium.

The approximate adiabatic index $\hat{\gamma}$ adopted above combined with the exact Blandford-McKee solutions
imply that $n^\prime = 4 \Gamma n$ for arbitrary $\Gamma$, where $n^\prime$ ($n$) is the downstream (upstream) density.
The downstream energy-density is similarly given by $u^\prime = 4 \Gamma (\Gamma-1) n \mu m_p c^2$ where $\mu$ is the mean molecular-weight, and we adopt $\mu \simeq 0.62$ appropriate for solar composition.
This can be used to find the downstream electron temperature $\Theta \equiv kT_e^\prime / m_e c^2$ via Equation~(2) of \cite{Margalit&Quataert21} with
$\Theta_0 = \epsilon_T (\mu m_p / \mu_e m_e) (\Gamma-1)$. We assume the mean molecular weight per electron is
$\mu_e \simeq 1.18$, as expected for fully-ionized solar composition material.
The equation for $\Theta$ implies that a fraction $\epsilon_T$ of the post-shock energy density is carried by thermal electrons.
We similarly adopt the usual assumption that a post-shock magnetic field $B$ is amplified such that $B^2/8\pi = \epsilon_B u^\prime$,
and that a fraction $\epsilon_e 
\equiv \delta \epsilon_T$ of the post-shock energy density is carried by non-thermal electrons that are accelerated into a power-law distribution $dn/d\gamma_e \propto \gamma_e^{-p}$ in electron Lorentz factor $\gamma_e$
(note that $B$ is the magnetic field in the post-shock frame; we omit the prime notation for consistency with previous work).

Based on the results of kinetic simulations (e.g., \citealt{Sironi+13}) we adopt the following fiducial values for the microphysical parameters described above: $\epsilon_T = 0.4$, $\epsilon_e = 0.01$ ($\delta = 0.025$), $\epsilon_B = 0.1$, and $p=3$.
This implies that 40\% of the post-shock energy is carried by thermal electrons, 10\% by magnetic fields, and 1\% by non-thermal (power-law) accelerated electrons. The remaining $\sim$50\% of post-shock energy is carried by thermal and non-thermal (cosmic ray) ions.
Typical kinetic simulations suggest that $\sim$10\% of energy goes into cosmic-ray ions and therefore that thermal ions carry $\sim$40\% of the post-shock energy. This situation implies approximate post-shock equilibration between electrons and ions such that $T_e \sim T_{\rm ion}$.
Considering lower values of $\epsilon_T$ would relax this assumption and generally allow for $T_e < T_{\rm ion}$.
We discuss these assumptions further in \S\ref{sec:microphysics}.

Following \cite{Sari+98,Sari+99}, the flux density emitted by a ``spherical'' relativistic source is $F_\nu \sim \pi \Gamma R^2 I_\nu^\prime / D_{L}^2$, where $D_L$ is the luminosity distance to the source 
and $I_\nu^\prime$ is the specific intensity in the source frame. The latter satisfies the radiative transfer equation $dI_\nu^\prime/ds = j_\nu^\prime - \alpha_\nu^\prime I_\nu^\prime$ where $j_\nu^\prime$ and $\alpha_\nu^\prime$ are the local emissivity and absorption coefficients.
The flux density estimate above is correct as long as the the source is azimuthally uniform (``spherical'') within the region from which (potentially beamed) radiation reaches the observer, i.e. within an opening angle $\sim 1/\Gamma$ from the observer LOS. Therefore, even a collimated source can be considered spherical if it is expanding relativistically and its opening angle is $> 1/\Gamma$ \citep{Sari+99}.
Defining the isotropic equivalent luminosity of a source as $L_{\nu} \equiv 4\pi D_L^2 F_\nu$, we can write
\begin{equation}
\label{eq:Lnu_full}
    L_{\nu} \approx 
    4\pi^2 \Gamma R^2 \frac{\left\langle j^\prime_{\nu} \right\rangle}{\left\langle \alpha^\prime_{\nu} \right\rangle} \left( 1 - e^{- \left\langle \alpha^\prime_{\nu} \right\rangle \Delta R^\prime } \right)
    .
\end{equation}
Here $\Delta R^\prime \sim R/\Gamma$ is the width of the emitting region
measured in the source frame
(in the observer frame, $\Delta R \sim R/\Gamma^2$),
and we define an emission volume-filling factor $f$ following \cite{Chevalier98} such that 
\begin{equation}
\label{eq:DeltaR_prime}
    \Delta R^\prime \equiv 4fR/3\Gamma .
\end{equation}
From hydrodynamic considerations one expects that $\Delta R^\prime \lesssim R/4\Gamma$. 
Furthermore, requiring that the number of electrons in the circumstellar medium that have been swept up by the shock equals the number of electrons within the post-shock shell (assuming a constant density shell of emitting matter) leads to
%\begin{equation}
%    f \leq \frac{3}{16 (3-k)}
%\end{equation}
$f \leq {3}/{\left[ 16 (3-k) \right]}$
where $0 \leq k < 3$ is the slope of the circumstellar density profile, such that $n(r) \propto r^{-k}$. 
A larger value of $f$ would imply that the number of particles within the post-shock shell is larger than the number of swept-up particles, which is not possible. Smaller values of $f$ are permissible if one assumes that only a fraction of the total swept-up material contributes to synchrotron emission (and absorption).
For our estimates below we adopt $f = 3/16$ as a fiducial value. 

In Equation~(\ref{eq:Lnu_full}) above,
the LOS averaged emissivity and absorption coefficients, 
$\left\langle j^\prime_{\nu} \right\rangle$, $\left\langle \alpha^\prime_{\nu} \right\rangle$,
are given by \cite{Margalit&Quataert21} and include the contribution of both thermal and non-thermal (power-law) electrons. 
In the non-relativistic limit 
and with $f=3/4$, 
Equation~(\ref{eq:Lnu_full}) reduces to Equation~(21) of \cite{Margalit&Quataert21}. In cases where only power-law electrons contribute to observed emission, it is also consistent with the estimates of \cite{Chevalier98} in the non-relativistic limit,
and with \cite{Sari+98} in the ultra-relativistic optically-thin limit. 

As previously mentioned, in the relativistic case one must generally integrate the specific intensity over a region from which all emitted photons arrive at the same time \citep{Granot+99a,Granot+99b}. In Equation~(\ref{eq:Lnu_full}) we have instead adopted an effective LOS approximation and neglected this aspect for the sake of analytic tractability.
This is similar to the approach taken by \cite{Sari+98}.
The impact of this assumption will be investigated in future work.\footnote{
On a technical note---we have called the formalism above an `effective' LOS approach because, although it does rely on estimating the total flux based on the intensity along some representative ray (that is, emission from some representative region), this region is not precisely along the line-of-sight between the origin of the explosion the observer 
(otherwise the Doppler factor would be larger by a factor $1+\beta$). 
%If it were then, for example, the corresponding Doppler factor would be $(1+\beta)\Gamma$. 
This choice is equally valid given that, in any case, the approach is merely an approximation to the more complicated relativistic radiative transfer problem that must be solved. We chose this particular formalism because it coincides with the assumptions of \cite{Sari+98} for GRBs.
}

With the assumptions above, the properties of the shock and its resulting emission can be fully specified with only two physical variables (along with several dimensionless microphysical and geometric parameters). In this paper we choose these physical variables to be the proper-velocity of the shock and the ambient density.
We find it convenient to express the ambient (upstream) density $n$ at the location of the shock in terms of
an effective mass-loss parameter $\dot{M}/v_{\rm w}$, such that
\begin{equation}
\label{eq:n_of_Mdot}
    \dot{M}/v_{\rm w} \equiv 4\pi \mu m_p R^2 n(R) .
\end{equation}
For reasons that will later become apparent,
this parameterization is useful because the mass-loss parameter depends on the peak frequency and observed time only through the combination $\nu_{\rm pk}t$. 
Note that Equation~(\ref{eq:n_of_Mdot}) is merely a parameterization choice which replaces $n \leftrightarrow \dot{M}/v_{\rm w}$. In particular, our results are {\it not} limited to the restricted case of a steady-state wind medium, and can be applied to arbitrary density profiles.\footnote{
One caveat to this statement is that emission is assumed to be dominated by electrons that have been swept-up by the most recent shock-doubling time. This typically requires that most of the swept-up material must be accumulated at $r \sim R$, and for a power-law density gradient $n \propto r^{-k}$, implies that $k<3$.
}
The mass-loss parameter should therefore be viewed as only a proxy for the ambient density at the current shock location.

\section{The Spectral Energy Distribution}
\label{sec:SEDs}

\begin{figure*}
    \centering
    \includegraphics[width=0.99\textwidth]{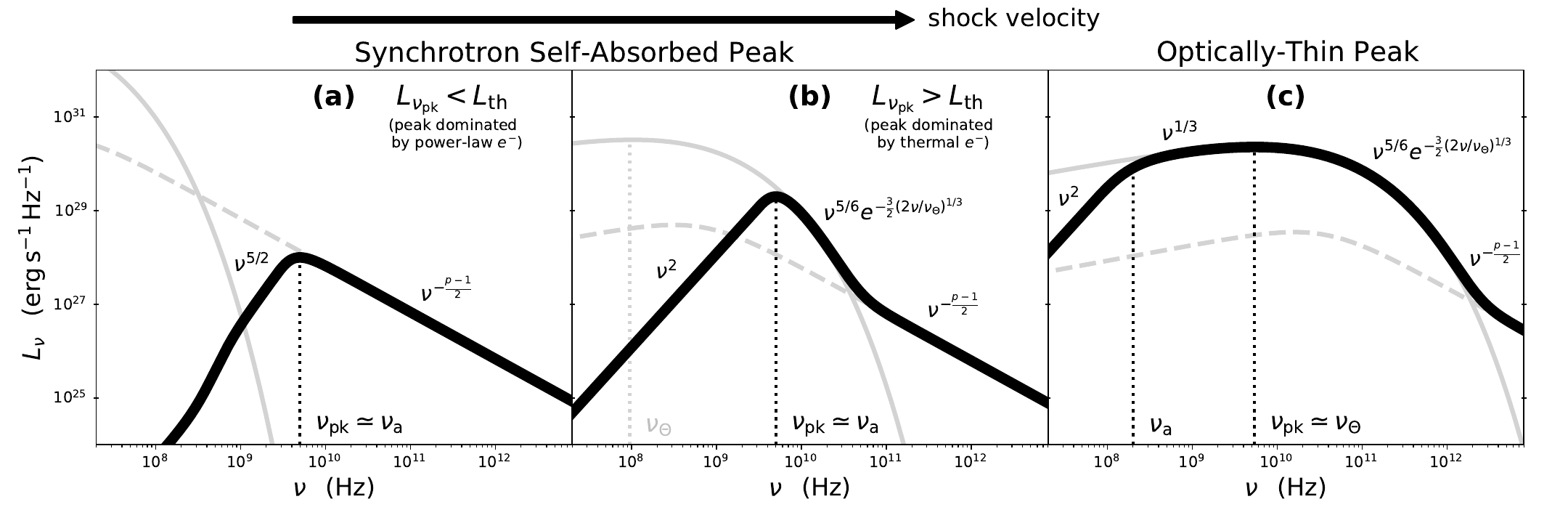}
    \caption{Representative spectral energy distributions (SEDs) for the three regimes considered in this work. The progression from left to right is a function (primarily) of the shock velocity.
    Solid and dashed grey curves show the emissivity contributed by the population of thermal and power-law electrons, respectively. Solid black curves show the resulting SED. The frequency scaling of $L_\nu$ is shown above each corresponding section of the SED.
    {\bf (a)} For $(\Gamma\beta)_{\rm sh} \approx 0.1$. The peak of the SED is set by synchrotron self-absorption (SSA) such that $L_{\nu_{\rm pk}}$ is obtained at the self-absorption frequency ($\nu_{\rm pk} \simeq \nu_{\rm a}$). At low shock velocities, the emissivity at $\nu_{\rm pk}$ is dominated by power-law electrons such that $L_{\nu_{\rm pk}} < L_{\rm th}$ (Equation~\ref{eq:L_th}). This regime corresponds to the `standard' model considered in the literature (e.g., \citealt{Chevalier98}) and is discussed here in \S\ref{sec:powerlaw}.
    {\bf (b)} For $(\Gamma\beta)_{\rm sh} \approx 0.6$. When the shock velocity is higher (typically mildly or trans-relativistic) then the SED peak is still governed by SSA, however emission at frequency $\nu_{\rm pk}$ is now dominated by thermal electrons. This is true when $L_{\nu_{\rm pk}} > L_{\rm th}$ (Equation~\ref{eq:L_th}). This regime implies a potentially steep spectral slope at frequencies slightly above $\nu_{\rm pk}$. At higher frequencies power-law electrons again dominate the emissivity and the spectral slope reverts the standard predictions (see \citealt{Margalit&Quataert21} for more details).
    {\bf (c)} For $(\Gamma\beta)_{\rm sh} \approx 3$. For ultra-relativistic shocks the SSA frequency falls below the peak of the emissivity ($\nu_{\rm a} < \nu_\Theta$) and the SED peaks at the latter frequency ($\nu_{\rm pk} \simeq \nu_\Theta$). Emission at $\nu_{\rm pk}$ is again dominated by thermal electrons, and this regime is discussed in \S\ref{sec:optically-thin}.
    }
    \label{fig:SEDs}
\end{figure*}

Following \cite{Chevalier98} we are interested in estimating physical properties of the shock in terms of the observed frequency $\nu_{\rm pk}$ and luminosity $L_{\nu_{\rm pk}}$ at which the SED peaks. 
We assume that electrons are slow cooling at this frequency and neglect extrinsic effects on the SED such as free-free absorption. 
As described in \S\ref{sec:model}, we extend the \cite{Chevalier98} model by considering both relativistic effects and the contribution of synchrotron emission from thermal electrons. The latter was shown to be important for mildly-relativistic shocks where $\beta_{\rm sh} \gtrsim 0.2$ \citep{Margalit&Quataert21}, while the former is clearly relevant when $(\Gamma\beta)_{\rm sh} \gtrsim 1$.

The unabsorbed SED is proportional to $j_\nu(\nu)$ and is intrinsically peaked at frequency $\nu_\Theta^\prime = 3 \Theta^2 e B / 4\pi m_e c$ \citep{Margalit&Quataert21} 
in the rest-frame of emitting matter, or $\nu_\Theta = \mathcal{D} \nu_\Theta^\prime$ in the observer frame. This corresponds to, 
%The unabsorbed SED is proportional to $j_\nu(\nu)$ and is intrinsically peaked at frequency $\nu_\Theta = 3 \Gamma \Theta^2 e B / 4\pi m_e c$ \citep{Margalit&Quataert21}, 
the frequency at which a typical thermal electron radiates most of its synchrotron power.
The observed SED however will be modified due to SSA and thus does not track the emissivity at frequencies below the SSA frequency, $\nu_{\rm a}$.
Two categories therefore emerge when considering the observed peak of the SED. In the first case, the intrinsic peak of the emissivity at $\nu \simeq \nu_\Theta$ is hidden due to SSA, and the observed peak of the SED occurs instead at the SSA frequency $\nu_{\rm a}$. This happens when $\nu_\Theta < \nu_{\rm a}$, that is---when the SSA optical depth at frequency $\nu_\Theta$ is $\tau_\Theta > 1$.
The second case occurs when $\tau_\Theta < 1$ and the SSA frequency falls below the thermal frequency, $\nu_{\rm a} < \nu_\Theta$. In this case the intrinsic peak of the emissivity is unabsorbed and corresponds to the observed peak of the SED ($\nu_{\rm pk} \simeq \nu_\Theta$).
The SSA optical depth is a strong function of shock velocity \citep{Margalit&Quataert21}, and we show in subsequent sections that the two scenarios described above are delineated by whether the velocity is smaller or larger than $\left(\Gamma\beta\right)_{\rm sh} \sim 1$.
The SSA SED can further be divided into two subcases depending on whether thermal electrons or power-law electrons dominate emission at frequency $\nu_{\rm pk}$. As discussed above, these two cases correspond to shock velocities above or below $\left(\Gamma\beta\right)_{\rm sh} \sim 0.3$, respectively.

We therefore have three cases to consider overall, delineated by increasing shock velocity. Each of these cases is characterized by a qualitatively different SED. Representative SEDs for these three regimes are illustrated in Figure~\ref{fig:SEDs}.
In the left panel (case a) the shock velocity is low and $\nu_{\rm pk} \simeq \nu_{\rm a}$ with emission dominated by power-law electrons. This is the `standard' case considered in the literature, corresponding to the original \cite{Chevalier98} analysis. The middle panel (case b) shows the SED for mildly-relativistic shocks where $\nu_{\rm pk} \simeq \nu_{\rm a}$ but emission at $\nu_{\rm pk}$ is dominated by thermal electrons. This regime was first discussed by \cite{Margalit&Quataert21}. Finally, the right panel (case c) shows the SED at even higher velocities (in particular, when $\Gamma\beta \gg 1$) when $\nu_{\rm a} < \nu_{\rm pk} \simeq \nu_\Theta$.
In the following sections we demonstrate how the shock properties can be inferred from observations of $\nu_{\rm pk}$ and $L_{\nu_{\rm pk}}$. In doing so, we separate our discussion based on the three regimes discussed above: \S\ref{sec:powerlaw} considers case (a); \S\ref{sec:thermal} treats case (b); and \S\ref{sec:optically-thin} focuses on case (c).

\section{
Synchrotron Self-Absorbed Peak: 
$\left(\Gamma\beta\right)_{\rm sh} \ll$--$\lesssim 1$
%$\left(\Gamma\beta\right)_{\rm sh} < \left(\Gamma\beta\right)_{\rm crit} \sim 1$
}
\label{sec:SSA}

\subsection{Power-law Electrons}
\label{sec:powerlaw}

\cite{Chevalier98} first pointed out the importance of SSA in shock-powered transients, and treated the case of a purely power-law distribution of emitting electrons. This regime corresponds to the SED illustrated in panel (a) of Figure~\ref{fig:SEDs}. We begin by briefly recapping these results for sake of completeness. All numerical values below adopt a fiducial $p=3$, but do not depend strongly on the power-law exponent.

In the limit $\left(\Gamma\beta\right)_{\rm sh} \ll 1$ and when power-law electrons dominate both the emissivity and absorption coefficients, the shock velocity can be expressed as
\begin{align}
\label{eq:betaGamma_pl}
    \left(\Gamma\beta\right)_{\rm sh}
    &\simeq
    %0.436 
    0.44 
    \left(\frac{\epsilon_e / \epsilon_B}{0.1}\right)^{-1/19} 
    \left(\frac{L_{\nu_{\rm pk}}}{10^{29}\,{\rm erg \,s}^{-1}\,{\rm Hz}^{-1}}\right)^{9/19} 
    \nonumber \\
    &\times
    \left(\frac{f}{3/16}\right)^{-1/19}
    \left(\frac{\nu_{\rm pk} \bar{t}}{5\,{\rm GHz} \times 100\,{\rm d}}\right)^{-1}
    ;~~L_{\nu_{\rm pk}} < L_{\rm th}
    .
\end{align}
The condition $L_{\nu_{\rm pk}} < L_{\rm th}$ is equivalent to the statement that power-law electrons dominate emission at frequency $\nu_{\rm pk} \simeq \nu_{\rm a}$, and is discussed in the following subsection (see Equation~\ref{eq:L_th}).
The mass-loss parameter can similarly be written as a function of the peak luminosity and frequency (e.g., \citealt{Chevalier&Fransson06}), giving
\begin{align}
\label{eq:Mdot_pl}
    \frac{\dot{M}}{v_{\rm w}} 
    %1.79
    &\simeq 1.8 \times 10^{-4} 
    \frac{M_\odot \,{\rm yr}^{-1}}{10^3\,{\rm km \,s}^{-1}} \,
    \epsilon_{B,-1}^{-11/19} \epsilon_{e,-2}^{-8/19}
    \nonumber \\
    &\times
    \left(\frac{f}{3/16}\right)^{-8/19}
    \left(\frac{L_{\nu_{\rm pk}}}{10^{29}\,{\rm erg \,s}^{-1}\,{\rm Hz}^{-1}}\right)^{-4/19}
    \nonumber \\
    &\times
    \left(\frac{\nu_{\rm pk} \bar{t}}{5\,{\rm GHz} \times 100\,{\rm d}}\right)^{2}
    ;~~L_{\nu_{\rm pk}} < L_{\rm th}
    .
\end{align}
Here and in the following we use standard notation whereby $\epsilon_{B,-1} \equiv \epsilon_B / 0.1$ and $\epsilon_{e,-2} \equiv \epsilon_e / 0.01$.

We note that the numerical prefactors in Equations~(\ref{eq:betaGamma_pl},\ref{eq:Mdot_pl}) differ slightly from the corresponding expressions in \cite{Chevalier98}.
Specifically, the shock velocity implied by Equation~(\ref{eq:betaGamma_pl}) is $\sim$30\% larger than would be inferred from Equation (13) in \cite{Chevalier98}.
This is due to two factors.
First, \cite{Chevalier98} expresses results in terms of a ``peak'' luminosity/frequency that is defined by the intersection of the optically-thin and optically-thick power-law limits, instead of using the true peak values as implied by Equation~(\ref{eq:Lnu_full}). This method overestimates the peak luminosity and underestimates peak frequency.\footnote{
More precisely, the ``peak'' frequency $\nu_*$ as defined in \cite{Chevalier98} is related to the true peak frequency $\nu_{\rm pk}$ via $\nu_* = \nu_{\rm pk} \tau_{\rm pk}^{2/(p+4)}$. 
The peak luminosity is similarly given by $L_{\nu_*} = L_{\nu_{\rm pk}} \tau_{\rm pk}^{5/(p+4)} / ( 1 - e^{-\tau_{\rm pk}} )$. 
The optical depth at peak $\tau_{\rm pk}$ can be found by numerically solving the equation $e^{-\tau_{\rm pk}} \left[ (p+4)\tau_{\rm pk} + 5 \right] = 5$ 
(e.g., \citealt{Ghisellini13} \S\,A.5). For $p=3$, $\tau_{\rm pk} \simeq 0.639$ and therefore $\nu_* \simeq 0.88 \nu_{\rm pk}$ and $L_{\nu_*} \simeq 1.54 L_{\nu_{\rm pk}}$.
}
Second, a small multiplicative difference between the two prescriptions arises due to different treatments of the electron pitch angle. \cite{Chevalier98} 
sets $\sin \theta = 1$ for simplicity whereas we assume a more realistic scenario where emitting electrons are randomly oriented (see \citealt{Margalit&Quataert21}).
Combining these effects, one can show that the above expressions are consistent with the \cite{Chevalier98} results if one replaces $L_{\nu_{\rm pk}} \to 2.02 L_{\nu_{\rm pk}}$ and $\nu_{\rm pk} \to 1.01 \nu_{\rm pk}$ in the equations of \cite{Chevalier98}.

\subsection{Thermal Electrons}
\label{sec:thermal}

We now turn to treating the second regime where thermal electrons dominate the emissivity and absorption coefficients at frequency $\nu_{\rm pk}$. This corresponds to the SED pictured in panel (b) of Figure~\ref{fig:SEDs}.

In this scenario, Equation~(\ref{eq:Lnu_full}) can be reduced to an expression for the peak luminosity
\begin{align}
\label{eq:Lpk_for_thermal_e}
    L_{\nu_{\rm pk}} 
    \approx
    &\frac{3\pi^2 \mu}{2 \mu_e} m_p c^2 \left(1-e^{-\tau_{\rm pk}}\right) \epsilon_T \left(\nu_{\rm pk} t\right)^2 
    \nonumber \\
    &\times
    \frac{\Gamma}{\Gamma+1} \left[1+\left(\Gamma\beta\right)_{\rm sh}^2\right] \left(\Gamma\beta\right)_{\rm sh}^4 
    ,
\end{align}
where $\tau_{\rm pk} \sim 1$ is the SSA optical depth at the peak frequency $\nu_{\rm pk} \simeq \nu_{\rm a}$.
Thermal electrons only dominate when the shock velocity, and therefore (according to the above equation) the peak luminosity, are high. Specifically, one can use Equation~(\ref{eq:Lpk_for_thermal_e}) to find that thermal electrons dominate when $L_{\nu_{\rm pk}} > L_{\rm th}$, where
\begin{align}
\label{eq:L_th}
    L_{\rm th} 
    &\approx
    1.7 \times 10^{28}\,{\rm erg \,s}^{-1}\,{\rm Hz}^{-1}\, \epsilon_{B,-1}^{-4/15} \epsilon_{e,-2}^{8/15} 
    \\ \nonumber
    &\times
    \left(\frac{\epsilon_T}{0.4}\right)^{-7/5} 
    \left(\frac{f}{3/16}\right)^{4/15}
    \left(\frac{\nu_{\rm pk} \bar{t}}{5\,{\rm GHz} \times 100\,{\rm d}}\right)^{34/15}
    .
\end{align}
This limit is obtained by requiring that $\nu_{\rm pk} = \nu_j$, where $\nu_j$ is the frequency below which the emissivity is dominated by thermal electrons \citep{Margalit&Quataert21}.
The dependence on $\epsilon_e$ here relies on the approximate relation $\nu_j/\nu_\Theta \propto (\epsilon_e/\epsilon_T)^{-0.25}$ \citep{Margalit&Quataert21},
and the fact that $L_{\rm th} \propto \epsilon_e^{4/15} (\nu_j/\nu_\Theta)^{-2(p+5)/15}$.

In the limit $\left(\Gamma\beta\right)_{\rm sh} \ll 1$, Equation~(\ref{eq:Lpk_for_thermal_e}) implies that the shock velocity is
\begin{align}
\label{eq:betaGamma_th}
    \left(\Gamma\beta\right)_{\rm sh}
    &\approx
    0.4 
    \left(\frac{\epsilon_T}{0.4}\right)^{-1/4}
    \left(\frac{L_{\nu_{\rm pk}}}{10^{29}\,{\rm erg \,s}^{-1}\,{\rm Hz}^{-1}}\right)^{1/4}
    \nonumber \\
    &\times 
    \left(\frac{\nu_{\rm pk} \bar{t}}{5\,{\rm GHz} \times 100\,{\rm d}}\right)^{-1/2}
    ;~~L_{\nu_{\rm pk}} > L_{\rm th}
\end{align}
when thermal electrons dominate,
an approximation that is typically accurate to within $\lesssim$15\% for $\left(\Gamma\beta\right)_{\rm sh} \leq 1$.
Note that, in contrast with the power-law electron case, here the shock velocity does not depend on $\epsilon_B$.

The mass-loss rate can similarly be expressed as a function of the peak luminosity and $\nu_{\rm pk}t$. In general, finding $\dot{M}/v_{\rm w}$ when thermal electrons dominate requires numerically solving a transcendental equation. However in the limit $\left(\Gamma\beta\right)_{\rm sh} \ll 1$, the following analytic approximation
holds to within a factor of a few,
\begin{align}
\label{eq:Mdot_th}
    \frac{\dot{M}}{v_{\rm w}} 
    &\sim 
    4
    \times 10^{-5} 
    \frac{M_\odot \,{\rm yr}^{-1}}{10^3\,{\rm km \,s}^{-1}}\,
    \epsilon_{B,-1}^{-2/3} \left(\frac{\epsilon_T}{0.4}\right)^{-11/12}
    \nonumber \\
    &\times
    \left(\frac{f}{3/16}\right)^{-1/3} 
    \left(\frac{L_{\nu_{\rm pk}}}{10^{29}\,{\rm erg \,s}^{-1}\,{\rm Hz}^{-1}}\right)^{-3/4} 
    \nonumber \\
    &\times
    \left(\frac{\nu_{\rm pk} \bar{t}}{5\,{\rm GHz} \times 100\,{\rm d}}\right)^{19/6}
    ;~~L_{\nu_{\rm pk}} > L_{\rm th}
    .
\end{align}
The above expression makes use of the approximate relation $\nu_{\rm pk}/\nu_\Theta \propto \tau_\Theta^{0.2}$ from \cite{Margalit&Quataert21},
where $\tau_\Theta$ is the SSA optical depth at frequency $\nu_\Theta$.

We can additionally use this approximate relation 
along with Equation (22) of \cite{Margalit&Quataert21}.
to express the spectral slope of the SED at $\nu \gg \nu_{\rm pk}$ as
\begin{align}
\label{eq:spectral_slope}
    &~~~~~~~~~~~\alpha(\nu)
    \equiv \left(\frac{d \ln L_\nu}{d \ln \nu}\right)
    \approx
    \frac{5}{6}
    - \tilde{\alpha}(\nu) ~;
    \\ \nonumber
    \tilde{\alpha}(\nu)
    &\sim
    3.2 
    \left(\frac{\nu}{\nu_{\rm pk}}\right)^{1/3}
    \epsilon_{B,-1}^{-1/18} \left(\frac{\epsilon_T}{0.4}\right)^{-13/72}
    \left(\frac{f}{3/16}\right)^{1/18} 
    \\ \nonumber
    &\times
    \left(\frac{L_{\nu_{\rm pk}}}{10^{29}\,{\rm erg \,s}^{-1}\,{\rm Hz}^{-1}}\right)^{-5/24} 
    \left(\frac{\nu_{\rm pk} \bar{t}}{5\,{\rm GHz} \times 100\,{\rm d}}\right)^{17/36}
    .
\end{align}
In principle, this provides a closure relation that can be tested against observed data. In practice however, this expression is only valid within a limited range of parameters where $\nu_\Theta,\nu_{\rm pk} \ll \nu \ll \nu_j$. These conditions are only satisfied if both $1 \ll 4\tilde{\alpha}(\nu)^3 \ll 380 \left(\delta/0.025\right)^{-1/4}$ and $\nu \gg \nu_{\rm pk}$. For the fiducial parameters above, these conditions are not met (the first condition is analogous to the requirement $L_{\rm th} \ll L_{\nu_{\rm pk}} \ll L_{\rm crit}$; Figure~\ref{fig:PhaseSpace_SSA} shows that this is rarely satisfied).

\subsection{The Phase-Space}

\begin{figure*}
    \centering
    \includegraphics[width=0.75\textwidth]{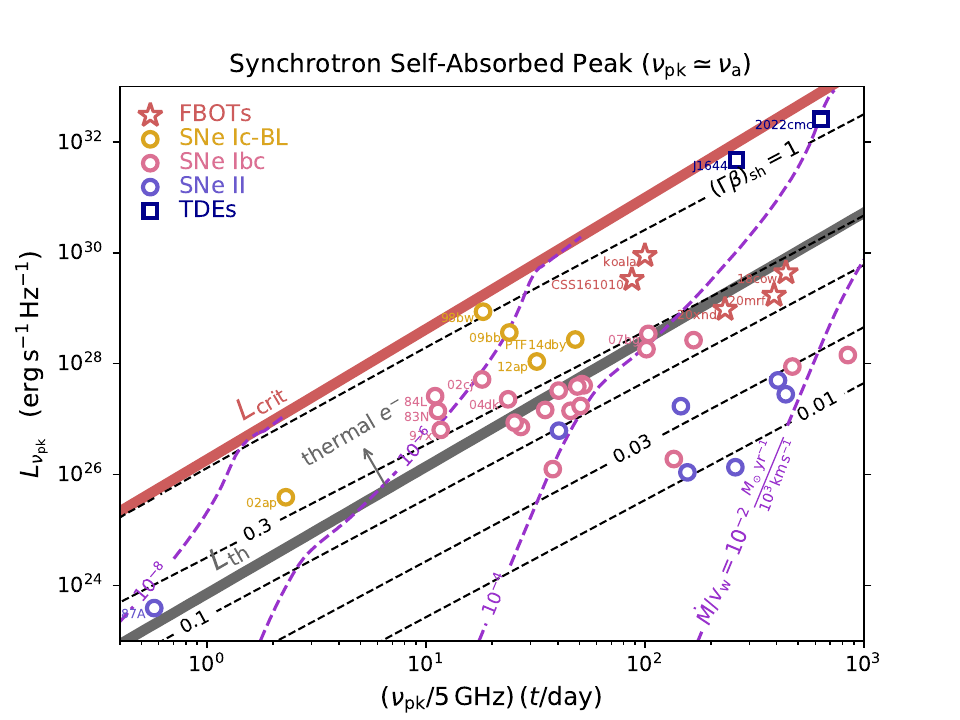}
    \caption{Phase-space of shock-powered synchrotron transients delineated in terms of the frequency $\nu_{\rm pk}$ and specific luminosity $L_{\nu_{\rm pk}}$ at which the spectral energy distribution (SED) peaks at time $t$: here, showing the regime where synchrotron self-absorption (SSA) sets these peak properties (such that $\nu_{\rm pk} \simeq \nu_{\rm a}$).
    Dashed black lines show contours of constant shock four-velocity $\left(\Gamma\beta\right)_{\rm sh}$, and dashed purple curves show contours along which the mass-loss parameter $\dot{M}/v_{\rm w}$ is constant (this is related to the density ahead of the shock through Equations~\ref{eq:R},\ref{eq:n_of_Mdot}).
    Below the solid grey curve, $L_{\nu_{\rm pk}} < L_{\rm th}$ (Equation~\ref{eq:L_th}) and power-law electrons dominate the emissivity at frequency $\nu_{\rm pk}$ (\S\ref{sec:powerlaw}). In this case, and when $\left(\Gamma\beta\right)_{\rm sh} \ll 1$, the results are consistent with the original analysis of \cite{Chevalier98}, and the SED takes the form shown in Figure~\ref{fig:SEDs} panel (a).
    Alternatively, when $L_{\nu_{\rm pk}} \gtrsim L_{\rm th}$ then thermal post-shock electrons dominate the synchrotron emissivity at frequency $\nu_{\rm pk}$, and must be taken into account (\S\ref{sec:thermal}; \citealt{Margalit&Quataert21}).
    The velocity and mass-loss contours within this region differ from the power-law-electron results and imply lower densities and shock velocities than would have otherwise been inferred from the standard \cite{Chevalier98} relations. The radio SED within this region is also qualitatively different and follows that shown in Figure~\ref{fig:SEDs} panel (b).
    The red curve labeled $L_{\rm crit}$ is an upper limit on the peak luminosity that we discuss in \S\ref{sec:optically-thin} 
    (see also Equation~\ref{eq:Appendix_Lcrit_approx}). 
    This curve corresponds to a SSA optical depth of $\tau_\Theta \sim 1$ at frequency $\nu_\Theta$, where the unabsorbed SED peaks, and is satisfied for shocks with $\left(\Gamma\beta\right)_{\rm sh} = \left(\Gamma\beta\right)_{\rm sh,crit} \sim 1$. The SED of higher velocity shocks is optically-thin at peak, with $\nu_{\rm pk} \simeq \nu_\Theta$. This situation is described by Figure~\ref{fig:PhaseSpace_OpticallyThin}.
    Here we have adopted fiducial values: $\epsilon_B = 0.1$, $\epsilon_T = 0.4$, $\epsilon_e=0.01$, 
    $p=3$, $f=3/16$, and $\ell_{\rm dec}=1$.
    }
    \label{fig:PhaseSpace_SSA}
\end{figure*}

Figure~\ref{fig:PhaseSpace_SSA} shows the phase-space of shock-powered SSA transients as a function of the frequency and specific luminosity at which the SED peaks at time $t$ (all measured in the observer frame). 
It accounts for relativistic effects as well as the contribution of thermal electrons---elements that are crucially important when considering high-velocity shocks.
A plot of this kind was first introduced by \cite{Chevalier98} for non-relativistic shocks (without thermal electrons),
and is widely used in analyses of radio transients. 
Its power derives from the fact that it allows one to infer and compare physical properties of different sources as a function of purely observational parameters.
In particular, the shock velocity $\left(\Gamma\beta\right)_{\rm sh}$ can be inferred from Figure~\ref{fig:PhaseSpace_SSA} using the dashed-black contours, while the upstream density (parameterized via the mass-loss parameter $\dot{M}/v_{\rm w}$; Equation~\ref{eq:n_of_Mdot}) can be found using the dashed-purple contours.

At low velocities, relativistic corrections and the effect of thermal electrons are negligible, and 
Figure~\ref{fig:PhaseSpace_SSA} is consistent with the standard \cite{Chevalier98} plot 
(although with minor quantitative corrections; see \ref{sec:powerlaw}).
However events that fall above $L_{\nu_{\rm pk}} > L_{\rm th}$ (thick grey curve; Equation~\ref{eq:L_th})
have their peak properties dominated by synchrotron emission from the population of thermal electrons in the post-shock region.
This modifies the \cite{Chevalier98} relationships and is most readily seen as kinks in the (purple) density curves, and an increased spacing of the shock-velocity contours (black curves).

Observed peak properties of various transients are also shown in Figure~\ref{fig:PhaseSpace_SSA}. Red stars denote FBOTs, circles denote different types of SNe, and squares show jetted TDEs. A subset of these events are identified by name, adjacent to each marker.
A growing number of events are found within the mildly-relativistic region where $L_{\nu_{\rm pk}} > L_{\rm th}$ and where our current relativistic model must be considered.
In particular, we find that using the non-relativistic \cite{Chevalier98} relations over-estimates both the velocity and density of such events (compare Equations~\ref{eq:betaGamma_th},\ref{eq:Mdot_th} with \ref{eq:betaGamma_pl},\ref{eq:Mdot_pl}), and we discuss some implications of this fact in \S\ref{sec:implications}.

The solid red curve in Figure~\ref{fig:PhaseSpace_SSA} shows an upper limit on the peak luminosity that we derive in the following section. It is interesting to note that, despite spanning many orders of magnitude in both $L_{\nu_{\rm pk}}$ and $\nu_{\rm pk}t$, not a single event exceeds the theoretical upper limit.

\section{
Optically-Thin Peak:
$\left(\Gamma\beta\right)_{\rm sh} \gg$--$\gtrsim 1$
}
\label{sec:optically-thin}

In the previous section we treated the case where the SED peak is set by SSA. However, for sufficiently high-velocity shocks the peak is instead determined by the intrinsic peak of the emissivity and $\nu_{\rm pk} \simeq \nu_\Theta$. This occurs once the SSA optical depth at frequency $\nu_\Theta$ drops below unity such that $\nu_{\rm a} < \nu_\Theta$ and $L_{\nu_{\rm a}} < L_{\nu_\Theta} \simeq L_{\nu_{\rm pk}}$.
The characteristic SED in this case is illustrated in Figure~\ref{fig:SEDs} panel (c).
In this scenario the peak luminosity 
is described by optically thin emission from thermal electrons 
whose characteristic Lorentz factor is $\gamma_e \sim 2\Theta$,
and can be written from Equation~(\ref{eq:Lnu_full}) as
\begin{equation}
\label{eq:Lpk_opticallythin}
    L_{\nu_{\rm pk}} 
    \overset{\left(\Gamma\beta\right)_{\rm sh} \gg 1}{\approx} 
    \frac{4 \pi^3 m_e c^2 f}{3 \epsilon_B \epsilon_T^6} \left(\frac{2 \sqrt{3} \mu_e m_e}{\mu m_p}\right)^7
    \frac{ \left(\nu_{\rm pk} \bar{t}\right)^3 }{ \left(\Gamma\beta\right)_{\rm sh}^{4} }
    .
\end{equation}
This equation is valid in the relativistic limit $\left(\Gamma\beta\right)_{\rm sh} \gg 1$, and can be used to obtain an explicit expression for the proper-velocity of the shock
\begin{align}
\label{eq:betaGamma_opticallythin}
    \left(\Gamma\beta\right)_{\rm sh}
    &\overset{\left(\Gamma\beta\right)_{\rm sh} \gg 1}{\approx} 
    3.1 \,
    \epsilon_{B,-1}^{-1/4}
    \left(\frac{\epsilon_T}{0.4}\right)^{-3/2}
    \left(\frac{f}{3/16}\right)^{1/4}
    \\ \nonumber
    &\times 
    \left(\frac{L_{\nu_{\rm pk}}}{10^{29}\,{\rm erg \,s}^{-1}\,{\rm Hz}^{-1}}\right)^{-1/4}
    \left(\frac{\nu_{\rm pk} \bar{t}}{5\,{\rm GHz} \times 100\,{\rm d}}\right)^{3/4}
\end{align}
in the optically-thin regime.
Using the fact that $\nu_{\rm pk} \simeq \nu_\Theta$ and the definition of $\nu_\Theta$ allows us to similarly express the mass-loss parameter in this regime as
\begin{align}
\label{eq:Mdot_opticallythin}
    \frac{\dot{M}}{v_{\rm w}} 
    &\overset{\left(\Gamma\beta\right)_{\rm sh} \gg 1}{\approx} 
    1.2
    \times 10^{-7} 
    \frac{M_\odot \,{\rm yr}^{-1}}{10^3\,{\rm km \,s}^{-1}}\,
    \left(\frac{\epsilon_T}{0.4}\right)^{2}
    \left(\frac{f}{3/16}\right)^{-1} 
    \nonumber \\
    &\times
    \left(\frac{L_{\nu_{\rm pk}}}{10^{29}\,{\rm erg \,s}^{-1}\,{\rm Hz}^{-1}}\right)
    \left(\frac{\nu_{\rm pk} \bar{t}}{5\,{\rm GHz} \times 100\,{\rm d}}\right)^{-1}
    .
\end{align}

These expressions are relevant for ultra-relativistic shocks such as in GRBs. Emission by thermal electrons in GRBs has been investigated by several studies in recent years \citep{Giannios&Spitkovsky09,Warren+17,Ressler&Laskar17,Warren+18,Samuelsson+20,Warren+22}. Our present results add to this growing literature by providing analytic expressions that allow one to infer the shock properties from the (thermal-electron-dominated) SED peak, if a clear peak in the SED is detected.
In practice however there are very few GRBs where a clear spectral peak in the broadband SED is observed. 
Additionally, the expressions derived here are only correct before the jet-break time (because of our assumption of spherical symmetry) and in the slow-cooling regime. These aspects limit the range of applicability of these results to GRB afterglows. An extension of this model to include these effects would be straightforward, but is outside the scope of our present work.

The ultra-relativistic results (Equations~\ref{eq:betaGamma_opticallythin},\ref{eq:Mdot_opticallythin}) may also be relevant in other astrophysical contexts apart from GRBs. For example the `synchrotron-maser' model for fast radio bursts (FRBs) invokes magnetized ultra-relativistic shocks as a mechanism for powering FRBs. A prediction of this model is that FRBs will be accompanied by high-energy synchrotron emission from thermal electrons behind the shock \citep{Lyubarsky14,Beloborodov17,Metzger+19,Beloborodov20,Margalit+20a,Margalit+20b}. Our present analysis would be suitable for describing such emission, albeit only if emitting electrons are slow-cooling (in the FRB context it is more likely that emission is instead fast-cooling; e.g., \citealt{Metzger+19}). To date, only a single FRB has an observed high-frequency counterpart that may be attributed to this model (the Galactic FRB~20200428; \citealt{Margalit+20b}). However future Galactic or local-Universe FRBs may increase this sample. 

Finally, we note the strong sensitivity on the composition of the ambient medium in the ultra-relativistic regime, namely $L_{\nu_{\rm pk}} \propto (\mu_e / \mu)^7$. For our assumed solar composition this amounts to a factor of $(\mu_e / \mu)^7 \approx 90$. Throughout much of the literature the composition is assumed to be pure H and the mean molecular weights are implicitly taken to be $\mu = \mu_e = 1$. Equation~(\ref{eq:Lpk_opticallythin}) demonstrates that, at least for peak emission in the ultra-relativistic limit, even a modest change to this assumption can cause a dramatic change to the luminosity (though note that this is  degenerate with $\epsilon_T$, which may itself depend on composition).

\subsection{A Limit on $L_{\nu_{\rm pk}}$}

An important consequence arises from Equations~(\ref{eq:Lpk_for_thermal_e},\ref{eq:Lpk_opticallythin}): 
while $L_{\nu_{\rm pk}}$ increases as a function of 
$\left(\Gamma\beta\right)_{\rm sh}$
when the shock is sub-relativistic (Equation~\ref{eq:Lpk_for_thermal_e}), it decreases with velocity when the shock is ultra-relativistic (Equation~\ref{eq:Lpk_opticallythin}).
This implies that, for fixed $\nu_{\rm pk} t$, there is a maximum attainable peak luminosity 
such that $L_{\nu_{\rm pk}} \leq L_{\rm crit}$. From the above argument it is also clear that this maximum is realized when $\left(\Gamma\beta\right)_{\rm sh} \sim 1$.
More precisely, there is a critical shock velocity $\left(\Gamma\beta\right)_{\rm sh, crit}$ at which $L_{\nu_{\rm pk}} = L_{\rm crit}$. Shocks with $\left(\Gamma\beta\right)_{\rm sh} > \left(\Gamma\beta\right)_{\rm sh, crit}$ are characterized by an optically-thin peak (case C in Figure~\ref{fig:SEDs}), while lower velocity shocks are SSA at peak (cases A and B in Figure~\ref{fig:SEDs}). Expressions for $\left(\Gamma\beta\right)_{\rm sh, crit}$ are derived in Appendix~\ref{sec:Appendix_Lcrit} (see Equations~\ref{eq:Appendix_bG_crit},\ref{eq:Appendix_bG_crit_approx}).

A crude estimate of the maximum peak luminosity $L_{\rm crit}$ that is acceptable for the fiducial parameters considered here is
\begin{align}
\label{eq:L_crit}
    L_{\rm crit} 
    &\sim 
    10^{31}\,{\rm erg \,s}^{-1}\,{\rm Hz}^{-1}\, \epsilon_{B,-1}^{-4/15} \left(\frac{\epsilon_T}{0.4}\right)^{-13/15} 
    \nonumber \\
    &\times
    \left(\frac{f}{3/16}\right)^{4/15}
    \left(\frac{\nu_{\rm pk} \bar{t}}{5\,{\rm GHz} \times 100\,{\rm d}}\right)^{34/15}
    ,
\end{align}
however this expression can be wildly inaccurate if $\epsilon_B \ll 0.1$, $\epsilon_T \ll 0.4$, or for extreme ranges of $\nu_{\rm pk} \bar{t}$ (even for our fiducial parameters, it is only accurate to within a factor of a few).
In Appendix~\ref{sec:Appendix_Lcrit} and Equations~(\ref{eq:Appendix_Lcrit},\ref{eq:Appendix_Lcrit_approx}) we derive more accurate expressions for $L_{\rm crit}$ that should generally be used instead of Equation~(\ref{eq:L_crit}).

\begin{figure}
    \centering
    \includegraphics[width=0.48\textwidth]{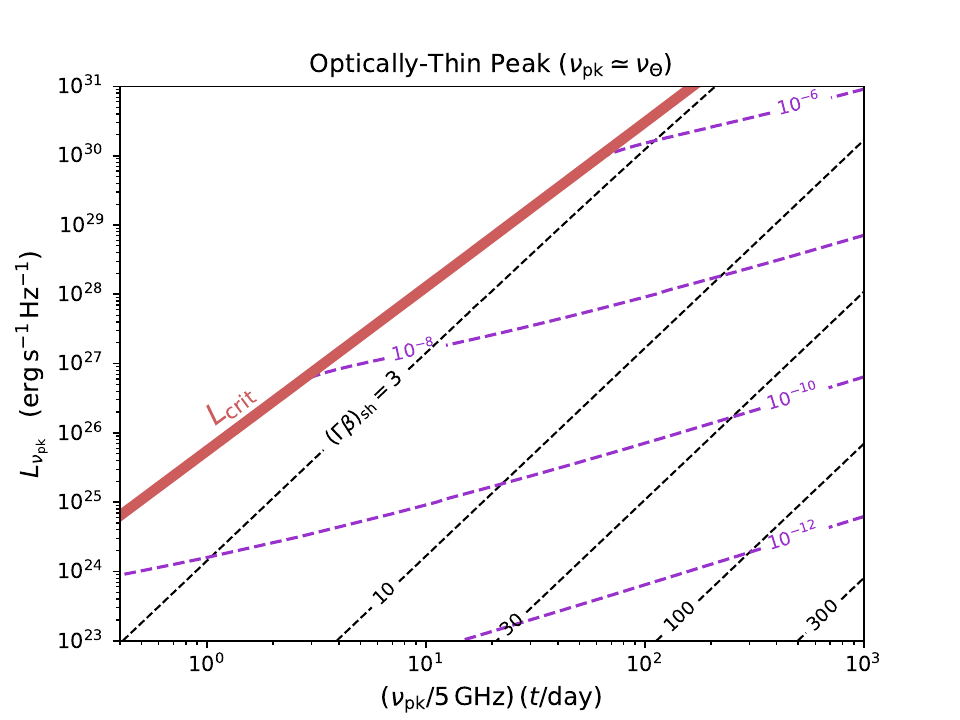}
    \caption{
    Same as Figure~\ref{fig:PhaseSpace_SSA}, but for the case where the SED peak is governed by optically-thin emission rather than SSA (that is, when $\nu_{\rm a} < \nu_\Theta \simeq \nu_{\rm pk}$; see Figure~\ref{fig:SEDs} panel c). This regime corresponds to $(\Gamma\beta)_{\rm sh} > (\Gamma\beta)_{\rm sh,crit} \sim 1$ and is relevant for relativistic shocks. See \S\ref{sec:optically-thin} for further details.
    }
    \label{fig:PhaseSpace_OpticallyThin}
\end{figure}

Figure~\ref{fig:PhaseSpace_OpticallyThin} shows the parameter-space of relativistic shocks, where the SED peak is set by optically-thin synchrotron emission from thermal electrons. Similar to Figure~(\ref{fig:PhaseSpace_SSA}), black dashed curves show contours of constant shock four-velocity while purple dashed curves show contours of constant mass-loss parameter. In this optically-thin regime the shock velocity increases as one moves rightwards on the diagram (towards large $\nu_{\rm pk} t$). This is the direct opposite of the situation in the SSA (sub-relativistic) regime shown in Figure~\ref{fig:PhaseSpace_SSA}.
The limiting peak luminosity $L_{\rm crit}$ can be understood by comparing these two figures. 
Consider a non-relativistic shock with $(\Gamma\beta)_{\rm sh} \ll 1$, as described by Figure~\ref{fig:PhaseSpace_SSA}. Now consider the change in $\nu_{\rm pk}$ and $L_{\nu_{\rm pk}}$ as one increases the shock velocity while keeping the mass-loss rate $\dot{M}/v_{\rm w}$ fixed. This is akin to moving upwards along a dashed-purple contour---towards higher $L_{\nu_{\rm pk}}$. Eventually, the peak luminosity hits the critical $L_{\rm crit}$ curve along which $\nu_{\rm pk} \approx \nu_{\rm a} \approx \nu_\Theta$. As one increases the shock velocity further, $\nu_{\rm a} < \nu_{\rm pk} \approx \nu_\Theta$ and the SED peak becomes governed by optically-thin emission. This situation is described by Figure~\ref{fig:PhaseSpace_OpticallyThin}. As can be seen, curves of constant mass-loss-rate undergo an inflection with respect to Figure~\ref{fig:PhaseSpace_SSA} such that---following the same mass-loss contour and increasing the velocity now causes one to move rightwards in the diagram, and away from the limiting $L_{\rm crit}$ curve. The limit $L_{\rm crit}$ can therefore be understood as a consequence of the fact that, in the optically-thin regime, the peak frequency $\nu_{\rm pk}$ increases with shock velocity much more rapidly than the peak luminosity does.

We conclude this section with an important note about the limiting luminosity. The luminosity $L_{\rm crit}$ is defined as the maximum {\it peak} luminosity attainable given observed values of $\nu_{\rm pk} t$. That is---$L_{\rm crit}$ is in general an upper limit on the specific luminosity at frequency $\nu_{\rm pk}$, where the SED peaks.
When $\nu_\Theta < \nu_{\rm a} \simeq \nu_{\rm pk}$ and SSA governs the peak flux (typically for sub-relativistic velocities) then $L_{\rm crit}$ is also a hard limit on the specific luminosity at {\it any} frequency (not just at $\nu_{\rm pk}$). This is because the SED at low frequencies $\nu < \nu_{\rm pk} \simeq \nu_{\rm a}$ is steep ($L_\nu \propto \nu^2$ or $\nu^{5/2}$ compared with $L_{\rm crit} \propto \nu^{34/15}$).
However when $\nu_{\rm a} < \nu_\Theta \simeq \nu_{\rm pk}$ (optically-thin peak) then $L_{\rm crit}$ is an upper limit only on the luminosity at the peak of the SED (at frequency $\nu_{\rm pk})$. 
The specific luminosity at low frequencies $\nu \ll \nu_{\rm pk}$ can exceed $L_{\rm crit}$
because the SED is relatively shallow between $\nu_{\rm pk}$ and the SSA frequency $\nu_{\rm a}$ ($L_\nu \propto \nu^{1/3}$; shallower than $L_{\rm crit} \propto \nu^{34/15}$).

\section{Disucssion and Implications}
\label{sec:implications}

A clear prediction of our model is that the radio SED of trans-relativistic and ultra-relativistic transients, where thermal electrons are important, should appear qualitatively different than the SED of non-relativistic shocks  (Figure~\ref{fig:SEDs}). Testing this hypothesis requires contemporaneous broadband observations with good frequency coverage. \cite{Ho+22} showed that the SEDs of the FBOTs AT2020xnd, CSS161010, and AT2018cow could be well-fit by a thermal electron model (e.g., see their Figure~11). Particularly noteworthy in this context is CSS161010 \citep{Coppejans+20}. It has one of the highest $L_{\nu_{\rm pk}}/L_{\rm th}$ ratios of any transient plotted in Figure~\ref{fig:PhaseSpace_SSA}, which implies that it should be deep within the thermal-electron-dominated regime. This event is also particularly well sampled with multi-frequency observations. As shown in \cite{Ho+22}, CSS161010 is remarkably well-fit by the thermal electron model, exhibiting the hallmark $L_\nu \propto \nu^2$ SSA slope expected in this regime (at least within a spherically-symmetric one-zone model) along with an apparently steepening optically-thin slope that is characteristic of the non-power-law electron distribution.

\begin{figure}
    \centering
    \includegraphics[width=0.48\textwidth]{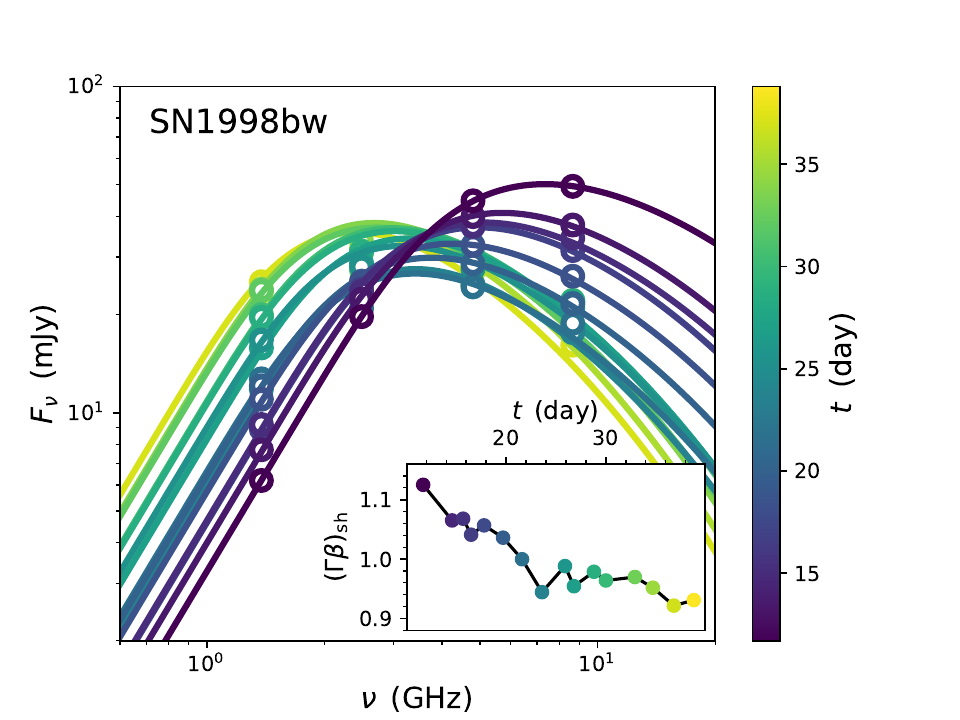}
    \caption{Multi-epoch fit to the radio SEDs of SN1998bw over the first 40 days (data from \citealt{Kulkarni+98}), assuming fiducial microphysical parameters (e.g., as in Figure~\ref{fig:PhaseSpace_SSA}). The best fit parameters imply that thermal electrons dominate emission in the observed frequency bands. It can be seen that the thermal synchrotron model provides excellent fits to the SEDs of 1998bw. The inset shows the inferred shock velocity that results from each epoch's SED fit. This is in line with the shock velocity shown in Figure~\ref{fig:PhaseSpace_SSA}.
    }
    \label{fig:98bw}
\end{figure}

To further test this model, we reinterpret data for the relativistic SN1998bw \citep{Kulkarni+98} using the model described in this paper. SN1998bw has one of the highest $L_{\nu_{\rm pk}}/L_{\rm th}$ ratios of any transient in our sample (see Figure~\ref{fig:PhaseSpace_SSA}), and is therefore a compelling test case for the thermal electron scenario. Figure~\ref{fig:98bw} shows multi-epoch fits to the radio SEDs of SN1998bw over the first $40$ days of its evolution. The inset shows the inferred shock velocity as a function of time, which is $(\Gamma\beta)_{\rm sh} \approx 1$ and decreasing with time, and the mass-loss rate is $\dot{M}/v_{\rm w} \sim 5 \times 10^{-7}\,M_\odot \, {\rm yr}^{-1} / 1000\,{\rm km \,s}^{-1}$. In this regime, thermal electrons dominate the observed emission. One can see that the thermal electron spectra provide an excellent fit to the data, further supporting this model.
In fact, in reviewing the literature on SN1998bw we became aware that such a model had already been proposed for this event due to the inconsistency of the observed SED with standard power-law model predictions \citep{Waxman&Loeb99}. Our renewed analysis lends further support to this idea.
Note that \cite{Waxman&Loeb99} adopted a mono-energetic electron distribution function for simplicity, while our treatment accounts for emission arising from a full thermal (Maxwell-J{\"u}ttner) distribution. The thermal distribution produces an SED that decays significantly less steeply at high frequencies 
(as $\propto \nu^{5/6} e^{-\frac{3}{2} \left( 2 \nu/\nu_\Theta \right)^{1/3}}$, instead of $\propto \nu^{1/2} e^{-\nu/\nu_\Theta}$)
and is therefore more broadband than that of a mono-energetic electron distribution.

We have also briefly tested this model against the SEDs of other high-velocity events shown in Figure~\ref{fig:PhaseSpace_SSA}. The jetted TDEs {\it Swift} J164449.3+573451 (J1644; \citealt{Bloom+11,Zauderer+11,Berger+12,Eftekhari+18}, and references therein) and AT2022cmc \citep{Andreoni+22,Pasham+23} can be well fit by a thermal-electron SED, and we will explore the implications of this in future work. We have also tested this model against data for several ``relativistic'' SNe, and find that in many cases the thermal-electron model can provide adequate fits. In a few cases, however, it seems that the thermal SED does not describe the data as well as a power-law electron model (one example is SN2009bb; \citealt{Soderberg+10}). One possibility is that the microphysical parameters in such events differ from our fiducial values (e.g., $\epsilon_B \ll 0.1$), such that $L_{\rm th} \gtrsim L_{\nu_{\rm pk}}$. For SN2009bb this would require that $\epsilon_{e,-2}^{-8/15} \epsilon_{B,-1}^{4/15} (\epsilon_T/0.4)^{7/5} \ell_{\rm dec}^{-34/15} \lesssim 0.03$. In general, the radio data for most SNe is sparse, so it is difficult to characterize the SED with great detail. As a byproduct, it is also common in many cases that the physical properties of the shock are determined by the peak of the light-curve in a given band, rather than the peak of the SED. Although the two may coincide in certain cases, this need not be the case more generally. If the density profile does not follow a simple power-law, or if there is energy injection into the shock, the light-curve can have peculiar and/or non-monotonic behavior, and it's peak need not coincide with the SSA frequency passing through the band. 
We warn against using events without a peak frequency that is well-determined from the SED in such analyses, and exclude events of this kind from our sample.

\subsection{Energy}
\label{sec:energy}

A primary impetus for this work is the increasing rate at which trans-relativistic transients are discovered. These transients are typically modeled using the canonical non-relativistic framework of \cite{Chevalier98}. However, as shown in this work, both relativistic corrections and the contribution of thermal electrons are important when modeling trans-relativistic explosions. 
A major implication of our present work is that properly taking these effects into account changes the inferred physical parameters for these events.

In particular, both thermal electrons and relativistic corrections act to decrease the inferred shock velocity and mass-loss rate. A direct consequence is that the associated shock energy is reduced when including these effects.
This can be seen by expressing the post-shock energy $U = V B^2/8\pi \epsilon_B$ as a function of observed peak properties. Taking $V = 4\pi f R^3/3$ and 
$B \approx 3 \epsilon_B^{1/2} (\dot{M}/v_{\rm w})^{1/2} / 2 \bar{t}$ (Equation~\ref{eq:Appendix_B_general})
we find using Equations~(\ref{eq:betaGamma_pl},\ref{eq:Mdot_pl},\ref{eq:betaGamma_th},\ref{eq:Mdot_th}) that
\begin{align}
\label{eq:U_pl}
    U 
    &\approx
    1.5 \times 10^{50}\,{\rm erg}\,
    \epsilon_{B,-1}^{-8/19} \epsilon_{e,-2}^{-11/19}
    \left(\frac{f}{3/16}\right)^{8/19}
    \\ \nonumber
    &\times 
    \left(\frac{L_{\nu_{\rm pk}}}{10^{29}\,{\rm erg \,s}^{-1}\,{\rm Hz}^{-1}}\right)^{23/19}
    \left(\frac{\nu_{\rm pk}}{5\,{\rm GHz}}\right)^{-1}
    ;~~L_{\nu_{\rm pk}} < L_{\rm th}
\end{align}
when power-law electrons dominate, while
\begin{align}
\label{eq:U_th}
    U 
    &\sim
    3 \times 10^{49}\,{\rm erg}\,
    \epsilon_{B,-1}^{-2/3}
    \left(\frac{\epsilon_T}{0.4}\right)^{-5/3}
    \left(\frac{f}{3/16}\right)^{2/3}
    \\ \nonumber
    &\times 
    \left(\frac{\nu_{\rm pk}}{5\,{\rm GHz}}\right)^{5/3}
    \left(\frac{\bar{t}}{100\,{\rm d}}\right)^{8/3}
    ;~~L_{\nu_{\rm pk}} > L_{\rm th}
\end{align}
when thermal electrons are responsible for the observed emission at frequency $\nu_{\rm pk}$.
In the regime where it is appropriate ($L_{\nu_{\rm pk}} > L_{\rm th}$), Equation~(\ref{eq:U_th}) gives a lower value of the post-shock energy than would have otherwise been inferred using the standard \cite{Chevalier98} model (Equation~\ref{eq:U_pl}).
In particular, the ratio of these inferred energies can be expressed as 
$U[\text{thermal }e^{-}]/U[\text{power-law }e^{-}] \sim \left( L_{\nu_{\rm pk}} / L_{\rm th} \right)^{-23/19}$,
with only very weak dependence on 
other parameters.\footnote{
$U[\text{thermal }e^{-}]/U[\text{power-law }e^{-}]
\propto \left( \epsilon_B / f \nu_{\rm pk} \bar{t} \right)^{0.08} \epsilon_T^{0.03} \epsilon_e^{-0.07}$
}
This illustrates how the inferred energy drops when one accounts for thermal electrons in the regime $L_{\nu_{\rm pk}} \gg L_{\rm th}$ where they are important.

\begin{figure*}
    \centering
    \includegraphics[width=0.75\textwidth]{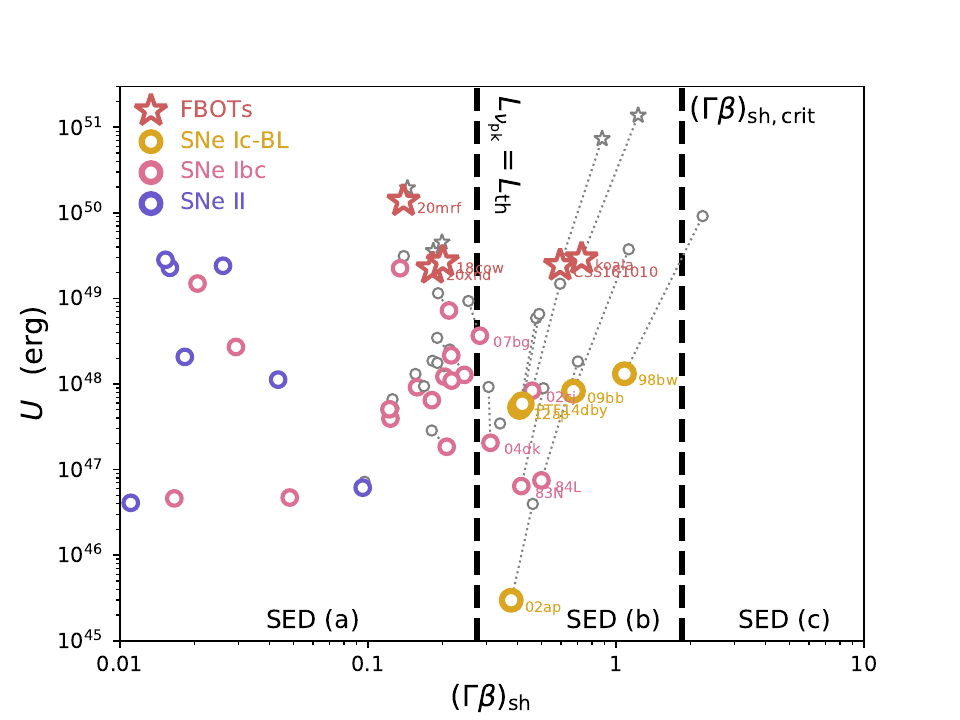}
    \caption{Post-shock energy versus shock velocity inferred from peak properties of synchrotron-powered transients. Colored symbols show results derived using the model described in this paper (different symbols and colors correspond to different classes of events). Connected grey points show the corresponding properties that would have been inferred using the commonly-adopted \cite{Chevalier98} model, which does not incorporate relativistic corrections or thermal electrons. These inferences overestimate the shock velocity and total energy, an effect that becomes pronounced for high-velocity explosions where $L_{\nu_{\rm pk}} > L_{\rm th}$ (to the right of the leftmost vertical black curve; see also Figure~\ref{fig:PhaseSpace_SSA} and Equation~\ref{eq:L_th}). The fact that the energetic constraints on some events can be reduced by more than an order of magnitude underpins the importance of considering relativistic effects when studying trans-relativistic transients.
    As one moves from low to high velocities the resulting SED changes in qualitative ways that provide a consistency check on the increasing importance of thermal electrons at high shock speeds. The regimes (a), (b), and (c), denoted here correspond to the three different SEDs illustrated in Figure~\ref{fig:SEDs}.
    The same fiducial microphysical and geometric parameters as in Figure~\ref{fig:PhaseSpace_SSA} are adopted here.
    }
    \label{fig:Energy}
\end{figure*}

Figure~\ref{fig:Energy} illustrates this point.
Colored symbols show the inferred energy and shock velocity for different transients as implied by our model. Red stars show the currently known sample of FBOTs, while differently colored open circles show various SNe. Each colored point is connected to a corresponding smaller grey symbol that shows the properties that would have been inferred for the same event if relativistic corrections and thermal electrons would have been neglected (in some cases the points overlap such that the grey symbols are hidden behind the colored ones).
Events that fall to the left of this plot (lower velocities) satisfy $L_{\nu_{\rm pk}} < L_{\rm th}$ such that their observed emission is dominated by power-law electrons and their SED will be similar to that shown in Figure~\ref{fig:SEDs} panel (a). Conversely, events that fall to the right of the leftmost vertical dashed-black curve have their peak emission dominated by thermal electrons ($L_{\nu_{\rm pk}} > L_{\rm th}$).\footnote{
The exact velocity at which $L_{\nu_{\rm pk}} = L_{\rm th}$ depends on $\nu_{\rm pk} t$ (albeit weakly; Equation~\ref{eq:L_th}) and will therefore vary for different events, see Figure~\ref{fig:PhaseSpace_SSA}. 
In Figure~\ref{fig:Energy} we plot this transition for a representative $\nu_{\rm pk} \bar{t} = 5\,{\rm GHz} \times 100\,{\rm day}$.
}
Figure~\ref{fig:Energy} shows that while the non-relativistic power-law-electron framework of \cite{Chevalier98} works well for events where the shock velocity is $\lesssim 0.2c$, it is inaccurate for trans-relativistic events with higher velocity. In particular, the shock velocity and energy are overestimated by this model (compare grey points with their associated colored symbols), the latter by as much as two orders of magnitude. This discrepancy becomes more acute as one moves towards higher shock velocities and further away from the $L_{\nu_{\rm pk}} = L_{\rm th}$ line.
This highlights the importance of considering relativistic corrections and thermal electrons when modeling such events.

Finally, we discuss the results of \cite{BarniolDuran+13} in light of our present work. \cite{BarniolDuran+13} generalized \cite{Chevalier98} and similar equipartition methods \citep[e.g.,][]{Scott&Readhead77} to the case of ultra-relativistic sources. In many ways this is similar to the objective of our present work. However, there are significant differences between the two analyses.
First, \cite{BarniolDuran+13} assume a purely power-law electron distribution and do not incorporate the effects of thermal electrons. As we have shown here, thermal electrons are critically important in relativistic shocks as they---and not the power-law electrons---determine the properties at the peak of the SED (see Figure~\ref{fig:SEDs} panel c).
Second, \cite{BarniolDuran+13} express their results in terms of a dimensionless parameter $\eta \equiv \min( \nu_{m}/\nu_{\rm a}, 1 )$ that is not specified in that work. Here $\nu_{m}$ is the characteristic (minimum) frequency of power-law electrons, and is analogous to $\nu_\Theta$ in the thermal electron scenario. Using the results of \cite{BarniolDuran+13} with $\eta$ treated as a constant $\sim \mathcal{O}(1)$ is therefore equivalent to the assumption that the peak of the SED occurs at $\nu_{\rm pk} \sim \nu_{\rm a}$. We have shown here instead that emission at the peak of the SED is typically optically-thin for ultra-relativistic shocks. To the extent that this conclusion applies also in the purely power-law electron scenario then one would generally expect $\nu_{\rm a} \ll \nu_{\rm pk} \simeq \nu_{\rm m}$, or $\eta \gg 1$
(examination of Table~3 in \citealt{Granot&Sari02} indicates that this is typically true for GRBs, for example).
The fact that $\eta \nsim 1$, and that the value of $\eta$ is not explicitly determined (it will, among other things, depend on $\Gamma$) therefore makes it difficult to compare the \cite{BarniolDuran+13} expressions with our present results.

\subsection{Microphysical Parameters}
\label{sec:microphysics}

The results of \S\ref{sec:energy} apply in the scenario where $\epsilon_e \ll \epsilon_T \lesssim 1$ (that is, $\delta \ll 1$). In the following subsection we discuss this assumption, which we consider to be physically well-motivated, albeit with significant uncertainties.
Before doing so, we note that
a minimum on the explosion energy is often estimated by instead adopting `equipartition' microphysical parameters, i.e. $\epsilon_e \sim \epsilon_B \sim \mathcal{O}(1)$. One can immediately see from Equation~(\ref{eq:U_pl}) that such values of $\epsilon_e$ and $\epsilon_B$ indeed minimize the inferred energy. 
In this scenario $\epsilon_e \sim \epsilon_T$ (i.e., $\delta \sim 1$), so that $L_{\rm th}$ is very large and there is little to no phase space in which thermal electrons dominate peak emission. That is, if power-law electrons are assumed to be in equipartition with magnetic fields and thermal electrons + ions, then $L_{\rm th} \sim L_{\rm crit}$ such that $L_{\nu_{\rm pk}}$ is almost always smaller than $L_{\rm th}$, and therefore Equation~(\ref{eq:U_pl})---and not~(\ref{eq:U_th})---is appropriate for estimating the energy. In other words, in the equipartition scenario thermal electrons are almost never relevant, even if they are present. This implies that the standard minimum-energy estimates, which neglect thermal electrons entirely, are still valid lower-limits on $U$.

The reduction in inferred energies when including thermal electrons that is shown in Figure~\ref{fig:Energy} and discussed in \S\ref{sec:energy} applies to the more physically-motivated case where $\epsilon_e \ll 1$ rather than equipartition. This would alleviate energetic constraints for more reasonable expected values of $\epsilon_e$ and $\epsilon_B$.
We also note that an equipartition argument holds true also in the case where thermal electrons are included and dominant over power-law electrons ($\epsilon_e \ll \epsilon_T$, $L_{\nu_{\rm pk}} > L_{\rm th}$). This can be derived from Equation~(\ref{eq:U_th}), which again shows that $U$ is minimized when $\epsilon_T \sim \epsilon_B \sim \mathcal{O}(1)$.

The details and efficiency of both non-thermal electron acceleration and electron-ion equilibration in collisionless shocks are in general not fully understood. This introduces significant uncertainties in the microphysical parameters $\epsilon_e$, $\epsilon_T$, and $\epsilon_B$.
As discussed in \S\ref{sec:model} and above, we choose $\epsilon_e=0.01$, $\epsilon_T=0.4$, and $\epsilon_B=0.1$ as fiducial values for these parameters.
Particle-in-cell (PIC) simulations generally support the idea that non-thermal particle acceleration occurs in collisionless shocks, however the efficiency of such acceleration (i.e. $\epsilon_e$) is subject to some uncertainty given numerical limitations on the run-time of these simulations. Some general trends are nevertheless noticeable and worth mentioning. 

First, the acceleration efficiency may depend on the strength and orientation of any pre-existing upstream magnetic field (to be differentiated from the downstream shock-amplified field). In the ultra-relativistic $\Gamma \gg 1$ regime, magnetized shocks (where the ratio of magnetic to kinetic flux crossing the shock front is $\gtrsim 1$) inhibit non-thermal particle acceleration and cause $\epsilon_e \to 0$ so long as the upstream magnetic field is oriented at an angle $\gtrsim 0.6 \,{\rm rad} /\Gamma \ll 1$ with respect to the shock normal (quasi-perpendicular shock; \citealt{Sironi&Spitkovsky09}). In the unmagnetized (or parallel shock) relativistic case, PIC simulations suggest that the non-thermal electron acceleration efficiency may be as high as $\epsilon_e \sim 0.1$ (e.g., \citealt{Sironi+13}). 
This is consistent with $\epsilon_e$ values that have been recently inferred for GRB radio afterglows (\citealt{Beniamini&vanderHorst17,Duncan+23}; although note that these results depend on modeling which neglects thermal electrons).
Lower velocity, non-relativistic, shocks have also been extensively studied using PIC simulations. In this case the efficiency of non-thermal particle acceleration depends more sensitively on the orientation of any pre-existing upstream magnetic field. However it is generally found that for non-relativistic shocks $\epsilon_e$ is smaller than for relativistic shocks. 
This is broadly consistent with findings regarding the cosmic-ray electron-to-proton ratio which imply that $\epsilon_e \ll \epsilon_p \sim 0.1$ \citep[e.g.,][]{Morlino&Caprioli12}.

Our present choice of fiducial $\epsilon_e = 10^{-2}$ is more in line with expectations for non-relativistic or mildly-relativistic shocks. If $\epsilon_e$ is larger in the relativistic regime then this may lead to a situation where $\epsilon_e \sim \epsilon_T$ for ultra-relativistic shocks. In this scenario, the effect of thermal electrons would be less significant because there would be little dynamic range (in frequency) where thermal electrons dominate over power-law ones. For example, in the optically-thin (relativistic) SED shown in the right-hand panel of Figure~\ref{fig:SEDs}, the transition to power-law-dominated emission (where the SED follows $L_\nu \propto nu^{-(p-1)/2}$) would happen much closer to $\nu_{\rm pk} = \nu_\Theta$ if $\epsilon_e \sim \epsilon_T$. This would inhibit the `thermal bump' and make the SED and light-curve look very-much like standard power-law models. We speculate that this may be one reason why there is, currently, no strong evidence supporting the presence of thermal electrons in GRB afterglows (although we emphasize that this deserves more attention in the literature).

The findings from PIC simulations that $\epsilon_e$ may depend on the shock velocity, increasing across the transition from non-relativistic to ultra-relativistic shocks, has important implications. Introducing a velocity-dependent $\epsilon_e(\Gamma\beta)$ would affect our results in various ways, as was first pointed out in \cite{Margalit&Quataert21}. This could be incorporated into our current framework, but is beyond the scope of this paper---particularly given the significant uncertainties and poor quantitative understanding of the relation $\epsilon_e(\Gamma\beta)$. It may be useful to revisit this in the future.

The detailed mechanisms and efficiency of electron-ion equilibration (and therefore the value of $\epsilon_T$) are also not fully understood at present.
Coulomb collisions are generally ineffective given their associated long timescales. Nevertheless, PIC simulations of shocks generically show that electrons gain energy and establish quasi-equilibrium with ions, even though the detailed plasma processes responsible for mediating this energy transfer are still debated.
Examining the limiting case where electrons do not exchange energy with ions at all places a strict lower limit of $\epsilon_T > \mu_e m_e / \mu m_p \simeq 10^{-3}$.
PIC simulations of relativistic shocks suggest instead that $\epsilon_T \sim O(1)$ (e.g., \citealt{Sironi+13,Groselj+24,Vanthieghem+24}). In the non-relativistic regime there is evidence that $\epsilon_T$ may be lower, and in particular data from SN remnants and from the solar wind suggest a strong dependence on the shock Mach number \citep[e.g.,][]{Sironi+13,Tran+20,Raymond+23}.

Given the current lack of a firm understanding of $\epsilon_T$ and its potential velocity dependence, we have adopted a constant $\epsilon_T = 0.4$ that is consistent with the naive scenario where electrons equilibrate with ions. 
This is also in line with recent work by \cite{Vanthieghem+24} which uses PIC simulations to study electron-ion equilibration and finds that $\epsilon_T \approx 0.3-0.4$ over a wide range of shock velocities $0.075 \lesssim \left(\Gamma\beta\right)_{\rm sh} \lesssim 3 \times 10^4$.\footnote{
\cite{Vanthieghem+24} define a parameter $\alpha$ such that $kT_e \equiv \alpha \left( \Gamma_{\rm sh} - 1 \right) \mu m_p c^2$. They find that $\alpha \approx 0.1$ nearly independent of the shock velocity. This parameter is related to our $\epsilon_T$ via $\epsilon_T = 8 \alpha / 3$ in the non-relativistic regime, and $\epsilon_T = 3\sqrt{2} \alpha$ for ultra-relativistic velocities.
}
Our assumption that $\epsilon_T = 0.4$ can be relaxed by inserting lower values of $\epsilon_T$ in any of the equations in this paper, and this will generally have a significant impact on many of the results.
%Flipping the uncertainty regarding these processes on its head, our present work on thermal-electron signatures can be viewed as a potential probe of the microphysics and may help constrain $\epsilon_T$.
A corollary of the uncertainty regarding $\epsilon_T$ is that our 
%present 
work makes $\epsilon_T$-dependent predictions that may help probe these microphysics and constrain $\epsilon_T$.

\section{Summary}
\label{sec:summary}

In this paper we have extended the seminal work of \cite{Chevalier98}, which considered SSA in non-relativistic shocks, to apply to shocks of arbitrary velocity. For the first time, we account for both relativistic corrections and the impact of post-shock thermal electrons with closed-form analytic expressions. These effects become significant when the shock proper-velocity exceeds $(\Gamma\beta)_{\rm sh} \gtrsim 0.2$ \citep{Margalit&Quataert21}, and are particularly important when considering trans-relativistic explosions.
We also treat the case where the SSA frequency falls below the intrinsic peak of the emissivity, and the observed peak emission is governed by optically-thin emission from thermal electrons. This regime 
occurs when $(\Gamma\beta)_{\rm sh} > (\Gamma\beta)_{\rm sh, crit} \sim 1$ (Equations~\ref{eq:Appendix_bG_crit},\ref{eq:Appendix_bG_crit_approx}),
and
is relevant for relativistic shocks.

Our main findings can be summarized as follows:
\begin{itemize}
    \item The physical properties of the shock can be determined from measurements of the peak frequency $\nu_{\rm pk}$  and spectral luminosity $L_{\nu_{\rm pk}}$ at which the synchrotron SED peaks.
    Technically, this relation is double-valued: observed $\left\{ \nu_{\rm pk}, L_{\nu_{\rm pk}} \right\}$ can be interpreted as either a SSA shock with $(\Gamma\beta)_{\rm sh} < (\Gamma\beta)_{\rm sh, crit}$, or a higher-velocity shock with $(\Gamma\beta)_{\rm sh} > (\Gamma\beta)_{\rm sh, crit}$ characterized by an optically-thin peak. 
    In practice however, it should be easy to distinguish between the two scenarios given the different SED morphology expected in either case (Figure~\ref{fig:SEDs}).
    \item For shocks governed by SSA:
    thermal electrons dominate the emissivity and absorption at frequency $\nu_{\rm pk}$ when $L_{\nu_{\rm pk}} > L_{\rm th}$ (Equation~\ref{eq:L_th}; solid grey curve in Figure~\ref{fig:PhaseSpace_SSA}). 
    For luminous events that fall within 
    this regime, the standard \cite{Chevalier98} relations (\ref{eq:betaGamma_pl},\ref{eq:Mdot_pl}) are no longer applicable. 
    Instead, the physical properties of the shock should be estimated from Equations~(\ref{eq:betaGamma_th},\ref{eq:Mdot_th}; see also Appendix~\ref{sec:Appendix})
    or using the numerical procedure outlined in Appendix~\ref{sec:Appendix_GeneralSolution} (implemented in code available at \href{https://github.com/bmargalit/thermal-synchrotron-v2}{https://github.com/bmargalit/thermal-synchrotron-v2}; see also \citealt{Margalit24_Code}).
    \item Accounting for thermal electrons 
    in the SSA regime
    reduces the velocities and mass-loss rates with respect to what would have otherwise been inferred using the standard \cite{Chevalier98} formalism. This leads to a significant reduction in the implied energy of trans-relativistic explosions, which may have previously been overestimated by as much as two orders of magnitude (Figure~\ref{fig:Energy}; Equations~\ref{eq:U_pl},\ref{eq:U_th}).
    \item A prediction of our model is that the radio SED will change qualitatively as a function of shock velocity. With increasing shock velocity, emission at $\nu_{\rm pk}$ transitions from being synchrotron self-absorbed and governed by power-law electrons, to a SSA peak dominated by thermal electrons, and eventually to an optically-thin peak (Figure~\ref{fig:SEDs}). This prediction can be tested with broadband observations of high-velocity events. As an example, we have modeled observations of SN1998bw and shown that the thermal electron model provides an excellent fit to the SEDs at $t \leq 40\,{\rm days}$ (Figure~\ref{fig:98bw}).
    The SEDs of several high-velocity FBOTs have also been shown to be well-described by the thermal electron model (\citealt{Ho+22}; their Figure~11), lending further support to this scenario.
    \item There is a maximum allowed peak luminosity $L_{\nu_{\rm pk}}$, such that $L_{\nu_{\rm pk}} \leq L_{\rm crit}$ 
    (Equation~\ref{eq:Appendix_Lcrit_approx}; see also Equation~\ref{eq:L_crit} and Figure~\ref{fig:PhaseSpace_SSA}).
    This is a consequence of the transition between optically-thin and optically-thick emission at the peak of the SED, and is a generic feature of synchrotron shock models.\footnote{
    The expression for $L_{\rm crit}$ is determined by properties of the thermal electron population (\S\ref{sec:optically-thin}; Appendix~\ref{sec:Appendix_Lcrit}). However, a qualitatively similar upper limit would apply also in the (unphysical) case where thermal electrons are neglected. This gives
    $L_{\rm crit} \sim 2 \times 10^{33}\,{\rm erg \,s}^{-1}\,{\rm Hz}^{-1} \, 
    \epsilon_{e,-2}^{-2} \epsilon_{B,-1}^{-23/55} 
    (\nu_{\rm pk} \bar{t} / 5{\rm GHz} \times 100{\rm d} )^{133/55}
    $.
    }
    While the specific luminosity at low frequencies $\nu \ll \nu_{\rm pk}$ can exceed this limit (that is, $L_\nu > L_{\rm crit}$ is possible) when peak emission is optically-thin, the {\it peak} luminosity should always satisfy $L_{\nu_{\rm pk}} \leq L_{\rm crit}$.
    It is interesting to note from Figure~\ref{fig:PhaseSpace_SSA} that no observed event falls above this limit (red curve), despite clear observational biases towards detecting more luminous events. We argue that the limit $L_{\rm crit}$ that we have derived here provides a natural explanation to this fact.
\end{itemize}

We have worked here under the assumption that electrons that dominate emission at frequency $\nu_{\rm pk}$ are slow cooling (that is, they do not lose appreciable energy throughout the duration of the observed emission). In Appendix~\ref{sec:Appendix_FastCooling} we derive the necessary conditions required for this assumption to be valid.
The result shows that the slow-cooling assumption is typically valid in the radio band, but does not always hold at higher frequencies.
In cases where electrons are fast cooling, our present results cannot be used (the shock velocity inferred from the expressions in this paper would underestimate the true velocity). The upper limit $L_{\nu_{\rm pk}} < L_{\rm crit}$ is likely to remain valid however, since the luminosity is lowered when electrons are fast-cooling.

We also neglect external propagation effects such as free-free absorption. If free-free absorption is important, it may modify the SED peak, in which case our formalism cannot be used. This is likely only important for events where the inferred ambient densities are very large. Such events generally fall towards the bottom-right corner of Figure~\ref{fig:PhaseSpace_SSA}. Explosions with large velocities---which are of primary interest here---are less-likely to be impacted by free-free absorption.

For the sake of clarity, we have also omitted the dependence on cosmological redshift $z$ from our expressions, in essence assuming that $z \approx 0$.
Incorporating cosmological corrections is straightforward however. One need only replace $t \to (1+z) t$ and $\nu_{\rm pk} \to \nu_{\rm pk} / (1+z)$ in any equation, and recall that 
$L_{\nu_{\rm pk}} = 4\pi D_{\rm lum}^2 F_{\nu_{\rm pk}} / (1+z)$
\citep[e.g.,][]{Hogg+99}, so that
\begin{equation}
    L_{\nu_{\rm pk}} 
    \approx \frac{10^{29}\,{\rm erg\,s}^{-1}\,{\rm Hz}^{-1} }{1+z}
    \left(\frac{F_{\nu_{\rm pk}}}{\rm mJy}\right) \left(\frac{D_{\rm lum}}{300\,{\rm Mpc}}\right)^2 .
\end{equation}
Here $D_{\rm lum}$ is the luminosity distance and $F_{\nu_{\rm pk}}$ the measured peak flux density.
Conveniently, many parameters are only dependent on time and frequency through the combination $\nu_{\rm pk} t$ which is invariant, and can therefore be calculated in any frame of reference.

Finally, we note that good spectral coverage is critical to properly diagnosing shocks within this framework. Our method relies on a reliable estimate of the frequency and luminosity at the peak of the SED, which can only be determined accurately with a well-sampled SED. In lieu of a well-sampled SED, the \cite{Chevalier98} methodology has often been used by substituting $\nu_{\rm pk} \leftrightarrow \nu$ and $t \leftrightarrow t_{\rm pk}$. 
That is, the method is sometimes used with an estimate of the time $t_{\rm pk}$ at which a single-band light-curve at some frequency $\nu$ peaks (as opposed to using the frequency $\nu_{\rm pk}$ at which the SED peaks).
This approach is only reasonable if the peak of the light-curve is set by the time at which $\nu_{\rm pk}$ passes through the observing band. This may indeed be true in certain cases, but is not generally guaranteed. The light-curve is influenced by the uncertain ambient density profile and shock expansion/deceleration history, neither of which need be well-behaved. Using the SED peak rather than the light-curve peak sidesteps these uncertainties, and is a more robust way of using the framework. This is true both in our present work, and for the original analysis of \cite{Chevalier98}.

The code developed in this analysis is publicly available 
%at \href{https://github.com/bmargalit/thermal-synchrotron-v2}{https://github.com/bmargalit/thermal-synchrotron-v2}, 
\citep{Margalit24_Code}
and may be particularly useful when the analytic approximations derived in the main text are not sufficiently accurate. This code can be imported as a python module and used to infer physical properties of observed transients, for example by fitting model-generated flux densities to real data. It may also be useful for generating plots similar to Figures (\ref{fig:PhaseSpace_SSA},\ref{fig:Energy}).

\begin{acknowledgements}
    B.M. thanks Nayana A.~J. for sharing a compilation of $\nu_{\rm pk}$ and $L_{\nu_{\rm pk}}$ for SNe which provided a basis for the sample of events shown in Figures~\ref{fig:PhaseSpace_SSA},\ref{fig:Energy}.
    This research benefited from interactions 
    %with ... (people)
    that were funded by the Gordon and Betty Moore Foundation through Grant GBMF5076.
\end{acknowledgements}

\begin{software}
    {\tt numpy} \citep{numpy}, {\tt scipy} \citep{scipy}, {\tt matplotlib} \citep{matplotlib},
    {\tt thermalsyn} \citep{Margalit24_Code}.
\end{software}


\bibliography{refs}{}
\bibliographystyle{aasjournal}

\appendix
\section{Expressions for Additional Physical Variables}
\label{sec:Appendix}

In the following, we use our main results to explicitly write solutions to other physical variables that describe the shock and that are of general interest. We begin by treating the case where SSA sets the peak of the SED, and find analytic solutions in the approximate limit where $\left(\Gamma\beta\right)_{\rm sh} \ll 1$. We then similarly treat the optically-thin case which is appropriate in the ultra-relativistic limit, $\left(\Gamma\beta\right)_{\rm sh} \gg 1$.

\subsection{SSA Peak}

When the peak of the SED is set by SSA (cases `a' and `b' in Figure~\ref{fig:SEDs}) we can use the results of \S\ref{sec:SSA}. In the limit $\left(\Gamma\beta\right)_{\rm sh} \ll 1$, the shock velocity is given by Equations~(\ref{eq:betaGamma_pl},\ref{eq:betaGamma_th}) depending on whether power-law or thermal electrons dominate emission at frequency $\nu_{\rm pk}$. Using these expressions along with Equation~(\ref{eq:R}), the shock velocity can be written as
\begin{equation}
\label{eq:Appendix_R}
    R \approx
    \begin{dcases}
    %1.1293 \times 10^{17}\,{\rm cm} \, 
    1.1 \times 10^{17}\,{\rm cm} \, 
    \left(\frac{\epsilon_e/\epsilon_B}{0.1}\right)^{-1/19} \left(\frac{f}{3/16}\right)^{-1/19}
    \left(\frac{L_{\nu_{\rm pk}}}{10^{29}\,{\rm erg \,s}^{-1}\,{\rm Hz}^{-1}}\right)^{9/19}
    \left(\frac{\nu_{\rm pk}}{5\,{\rm GHz}}\right)^{-1}
    &;~ L_{\nu_{\rm pk}} < L_{\rm th}
    \\
    %1.062 \times 10^{17}\,{\rm cm} \, 
    10^{17}\,{\rm cm} \, 
    \left(\frac{\epsilon_T}{0.4}\right)^{-1/4} \left(\frac{L_{\nu_{\rm pk}}}{10^{29}\,{\rm erg \,s}^{-1}\,{\rm Hz}^{-1}}\right)^{1/4}
    \left(\frac{\nu_{\rm pk}}{5\,{\rm GHz}}\right)^{-1/2}
    \left(\frac{\bar{t}}{100\,{\rm d}}\right)^{1/2}
    &;~ L_{\nu_{\rm pk}} > L_{\rm th}
    \end{dcases}
\end{equation}
where the top ($L_{\nu_{\rm pk}} < L_{\rm th}$) case corresponds to the power-law-dominated regime \citep{Chevalier98}, and the bottom case ($L_{\nu_{\rm pk}} > L_{\rm th}$) corresponds to the thermal-electron-dominated scenario \citep{Margalit&Quataert21}.
As discussed in the main text, $L_{\rm th}$ is the spectral luminosity that separates these two regimes (Equation~\ref{eq:L_th}; see also Figure~\ref{fig:PhaseSpace_SSA})

Next we find the post-shock magnetic field, which is related to the mass-loss rate via 
\begin{equation}
\label{eq:Appendix_B_general}
    B = \frac{c}{R} \sqrt{ 8 \epsilon_B (\dot{M}/v_{\rm w}) \Gamma (\Gamma-1) }
\end{equation}
In the limit $\left(\Gamma\beta\right)_{\rm sh} \ll 1$ this reduces to $B \approx 3 \epsilon_B^{1/2} (\dot{M}/v_{\rm w})^{1/2} / 2 \bar{t}$. Using this result along with the mass-loss rate as given by Equations~(\ref{eq:Mdot_pl},\ref{eq:Mdot_th}) we find the post-shock magnetic field
\begin{equation}
\label{eq:Appendix_B}
    B \approx
    \begin{dcases}
    %0.58328\,{\rm G}\,
    0.58\,{\rm G}\,
    \left(\frac{\epsilon_e/\epsilon_B}{0.1}\right)^{-4/19}
    \left(\frac{f}{3/16}\right)^{-4/19} 
    \left(\frac{L_{\nu_{\rm pk}}}{10^{29}\,{\rm erg \,s}^{-1}\,{\rm Hz}^{-1}}\right)^{-2/19} 
    \left(\frac{\nu_{\rm pk}}{5\,{\rm GHz}}\right)
    &;~ L_{\nu_{\rm pk}} < L_{\rm th}
    \\
    %0.28254\,{\rm G}\,
    0.3\,{\rm G}\,
    \epsilon_{B,-1}^{1/6} \left(\frac{\epsilon_T}{0.4}\right)^{-11/24}
    \left(\frac{f}{3/16}\right)^{-1/6} 
    \left(\frac{L_{\nu_{\rm pk}}}{10^{29}\,{\rm erg \,s}^{-1}\,{\rm Hz}^{-1}}\right)^{-3/8} 
    \left(\frac{\nu_{\rm pk}}{5\,{\rm GHz}}\right)^{19/12}
    \left(\frac{\bar{t}}{100\,{\rm d}}\right)^{7/12}
    &;~ L_{\nu_{\rm pk}} > L_{\rm th}
    \end{dcases}
\end{equation}
in the power-law and thermal electron cases, respectively.
The upstream density can similarly be derived using the mass-loss rate (Equations~\ref{eq:Mdot_pl},\ref{eq:Mdot_th}) along with the radius (Equation~\ref{eq:Appendix_R}) and Equation~(\ref{eq:n_of_Mdot}), giving
\begin{equation}
\label{eq:Appendix_n}
    n \approx
    \begin{dcases}
    %678.99\,{\rm cm}^{-3}\,
    680\,{\rm cm}^{-3}\,
    \epsilon_{B,-1}^{-13/19} \epsilon_{e,-2}^{-6/19} \left(\frac{f}{3/16}\right)^{-6/19}
    \left(\frac{L_{\nu_{\rm pk}}}{10^{29}\,{\rm erg \,s}^{-1}\,{\rm Hz}^{-1}}\right)^{-22/19}
    \left(\frac{\nu_{\rm pk}}{5\,{\rm GHz}}\right)^4
    \left(\frac{\bar{t}}{100\,{\rm d}}\right)^2
    &;~ L_{\nu_{\rm pk}} < L_{\rm th}
    \\
    %180.2\,{\rm cm}^{-3}\,
    200\,{\rm cm}^{-3}\,
    \epsilon_{B,-1}^{1/6} \left(\frac{\epsilon_T}{0.4}\right)^{-5/12}
    \left(\frac{f}{3/16}\right)^{-1/3} 
    \left(\frac{L_{\nu_{\rm pk}}}{10^{29}\,{\rm erg \,s}^{-1}\,{\rm Hz}^{-1}}\right)^{-5/4}
    \left(\frac{\nu_{\rm pk}}{5\,{\rm GHz}}\right)^{25/6}
    \left(\frac{\bar{t}}{100\,{\rm d}}\right)^{13/6}
    &;~ L_{\nu_{\rm pk}} > L_{\rm th}
    \end{dcases}
    .
\end{equation}
Finally, for the sake of completeness, we repeat the post-shock energy as given by Equations~(\ref{eq:U_th},\ref{eq:U_pl}),
\begin{equation}
\label{eq:Appendix_U}
    U \approx
    \begin{dcases}
    1.5 \times 10^{50}\,{\rm erg}\,
    \epsilon_{B,-1}^{-8/19} \epsilon_{e,-2}^{-11/19}
    \left(\frac{f}{3/16}\right)^{8/19}
    \left(\frac{L_{\nu_{\rm pk}}}{10^{29}\,{\rm erg \,s}^{-1}\,{\rm Hz}^{-1}}\right)^{23/19}
    \left(\frac{\nu_{\rm pk}}{5\,{\rm GHz}}\right)^{-1}
    &;~ L_{\nu_{\rm pk}} < L_{\rm th}
    \\
    3 \times 10^{49}\,{\rm erg}\,
    \epsilon_{B,-1}^{-2/3}
    \left(\frac{\epsilon_T}{0.4}\right)^{-5/3}
    \left(\frac{f}{3/16}\right)^{2/3}
    \left(\frac{\nu_{\rm pk}}{5\,{\rm GHz}}\right)^{5/3}
    \left(\frac{\bar{t}}{100\,{\rm d}}\right)^{8/3}
    &;~ L_{\nu_{\rm pk}} > L_{\rm th}
    \end{dcases}
    .
\end{equation}
Note that the bottom cases in Equations~(\ref{eq:Appendix_B},\ref{eq:Appendix_n},\ref{eq:Appendix_U}) depend on the mass-loss rate in the thermal-electron regime. The analytic expression for $\dot{M}/v_{\rm w}$ in this case (Equation~\ref{eq:Mdot_th}) is an approximate relationship that is only accurate to within a factor of a few. The magnetic field, density, and energy estimates are therefore also subject to these errors in the thermal electron scenario. 
These are a consequence of the transcendental nature of the equations that arise in the thermal electron case, which have no closed-form analytic solution. For analytic tractability we have adopted in the above equations an approximate solution to this transcendental equation (in particular, the equation relating the SSA frequency $\nu_{\rm a}$ to the optical depth at $\nu_\Theta$, $\tau_\Theta$) as described in \cite{Margalit&Quataert21}.
The code we provide can obtain exact solutions to these variables (under the assumptions of the model, see \S\ref{sec:model}).

\subsection{Optically-Thin Peak}

In the ultra-relativistic $\left(\Gamma\beta\right)_{\rm sh} \gg 1$ limit the peak of the SED is governed by optically-thin synchrotron emission dominated by thermal electrons and the results of \S\ref{sec:optically-thin} are applicable.
Using these results and Equation~(\ref{eq:R}) we can express the shock radius as
\begin{equation}
    %R \approx 2.44 \times 10^{18}\,{\rm cm} \, 
    R \approx 2.4 \times 10^{18}\,{\rm cm} \, 
    \epsilon_{B,-1}^{-1/2}
    \left(\frac{\epsilon_T}{0.4}\right)^{-3}
    \left(\frac{f}{3/16}\right)^{1/2}
    \left(\frac{L_{\nu_{\rm pk}}}{10^{29}\,{\rm erg \,s}^{-1}\,{\rm Hz}^{-1}}\right)^{-1/2}
    \left(\frac{\nu_{\rm pk}}{5\,{\rm GHz}}\right)^{3/2}
    \left(\frac{\bar{t}}{100\,{\rm d}}\right)^{5/2}
    .
\end{equation}
The post-shock magnetic field in the ultra-relativistic limit reduces to $B \approx 2 \epsilon_B^{1/2} (\dot{M}/v_{\rm w})^{1/2} / (\Gamma\beta)_{\rm sh} \bar{t}$ which, combined with Equations~(\ref{eq:betaGamma_opticallythin},\ref{eq:Mdot_opticallythin}) gives
\begin{equation}
    %B \approx 6.48\,{\rm mG}\,
    B \approx 6.5\,{\rm mG}\,
    \epsilon_{B,-1}^{3/4} \left(\frac{\epsilon_T}{0.4}\right)^{5/2}
    \left(\frac{f}{3/16}\right)^{-3/4} 
    \left(\frac{L_{\nu_{\rm pk}}}{10^{29}\,{\rm erg \,s}^{-1}\,{\rm Hz}^{-1}}\right)^{3/4}
    \left(\frac{\nu_{\rm pk}}{5\,{\rm GHz}}\right)^{-5/4}
    \left(\frac{\bar{t}}{100\,{\rm d}}\right)^{-9/4}
    .
\end{equation}
The upstream density is then (Equations~\ref{eq:n_of_Mdot},\ref{eq:Mdot_opticallythin})
\begin{equation}
    %n \approx 9.53 \times 10^{-4}\,{\rm cm}^{-3}\,
    n \approx 9.5 \times 10^{-4}\,{\rm cm}^{-3}\,
    \epsilon_{B,-1} \left(\frac{\epsilon_T}{0.4}\right)^{8}
    \left(\frac{f}{3/16}\right)^{-2} 
    \left(\frac{L_{\nu_{\rm pk}}}{10^{29}\,{\rm erg \,s}^{-1}\,{\rm Hz}^{-1}}\right)^2
    \left(\frac{\nu_{\rm pk}}{5\,{\rm GHz}}\right)^{-4}
    \left(\frac{\bar{t}}{100\,{\rm d}}\right)^{-6}
    .
\end{equation}
Finally, the post-shock energy $U = f B^2 R^3 / 6 \epsilon_B$ can be expressed as
\begin{equation}
    U \approx 1.9 \times 10^{50}\,{\rm erg}\,
    \epsilon_{B,-1}^{-1} \left(\frac{\epsilon_T}{0.4}\right)^{-4}
    \left(\frac{f}{3/16}\right) 
    \left(\frac{\nu_{\rm pk}}{5\,{\rm GHz}}\right)^2
    \left(\frac{\bar{t}}{100\,{\rm d}}\right)^3
    ,
\end{equation}
which does not depend on the peak luminosity.

\section{General Solution}
\label{sec:Appendix_GeneralSolution}

In the main text and the preceding Appendix we have derived analytic solutions that are appropriate in different regimes. These solutions are, in certain cases, only approximate, and are never exact at the transition between different regimes (e.g., near $L_{\nu_{\rm pk}} \sim L_{\rm th}$, or $L_{\nu_{\rm pk}} \sim L_{\rm crit}$). Here we outline a general methodology for solving for $(\Gamma\beta)_{\rm sh}$, $\dot{M}/v_{\rm w}$ given $\nu_{\rm pk}$, $L_{\nu_{\rm pk}}$, that is accurate for any set of parameters, but which generally requires a numerical procedure.

Following Equation~(\ref{eq:Lnu_full}), the specific luminosity at frequency $\nu$ (not necessarily the peak frequency) can be written in a general form as
\begin{align}
\label{eq:Appendix_L_general}
    L_\nu \left[ \nu ~;~ (\Gamma\beta)_{\rm sh},\dot{M}/v_{\rm w} \right] 
    = 4\pi^2 \Gamma R^2 \Delta R^\prime \langle j_\nu^\prime \rangle \frac{1-e^{-\tau(\nu)}}{\tau(\nu)}
    = 
    \tilde{L} \left[ x I^\prime(x) + \frac{8\pi}{\sqrt{3}} C_j(p) \delta \frac{g(\Theta)}{f(\Theta)} x^{-\frac{p-1}{2}} \right]
    \frac{1-e^{-\tau(x)}}{\tau(x)}
    ,
\end{align}
where
\begin{equation}
\label{eq:Appendix_tau}
    \tau(x) 
    \equiv \tau_\Theta \left[ \frac{I^\prime(x)}{x} + \frac{3^{3/2}}{\pi} C_\alpha(p) \delta \frac{g(\Theta)}{f(\Theta)} x^{-\frac{p+4}{2}} \right]
    ,
\end{equation}
is the optical depth at frequency $\nu$, $\delta = \epsilon_e/\epsilon_T$,
\begin{equation}
\label{eq:Appendix_x}
    x \equiv \frac{\nu}{\nu_\Theta}
    = \frac{\sqrt{2}\pi m_e c}{3e \epsilon_B^{1/2}} \left(\frac{\dot{M}}{v_{\rm w}}\right)^{-1/2} \frac{ \Gamma_{\rm sh}^2\beta_{\rm sh} }{\Gamma^{3/2} (\Gamma-1)^{1/2} \Theta^{2}} \nu \bar{t}
    %\overset{\Theta \gg 1}{\approx} \frac{3 \sqrt{2}\pi m_e c}{e \epsilon_B^{1/2} \epsilon_T^2} \left(\frac{\mu_e m_e}{\mu m_p}\right)^2 \left(\frac{\dot{M}}{v_{\rm w}}\right)^{-1/2} \frac{\Gamma_{\rm sh}^2 \beta_{\rm sh}}{\Gamma^{3/2} (\Gamma-1)^{5/2}} \, \nu \bar{t}
    ,
\end{equation}
is a normalized frequency, and
\begin{equation}
    \tilde{L} 
    \equiv \frac{4\sqrt{2} e^3 \mu_e \epsilon_B^{1/2} f}{\sqrt{3} \mu m_p m_e c} \left(\frac{\dot{M}}{v_{\rm w}}\right)^{3/2} f(\Theta) \Gamma^{3/2} (\Gamma-1)^{1/2}
    ,
\end{equation}
\begin{equation}
    \tau_\Theta \equiv \frac{\sqrt{2} e \mu_e f}{3^{5/2} \mu m_p c \epsilon_B^{1/2}} \left(\frac{\dot{M}}{v_{\rm w}}\right)^{1/2} \frac{f(\Theta)}{\Theta^5} \Gamma^{-1/2} (\Gamma-1)^{-1/2}
\end{equation}
are functions of the shock velocity and effective mass-loss rate.
In the above, $f(\Theta)$ and $g(\Theta)$ (not to be confused with the volume filling factor $f$) are functions of the normalized electron temperature and are given by Equations~(5,8) of \cite{Margalit&Quataert21}.\footnote{
Note that there was a typo in the expression for $g(\Theta)$ presented in Equation~(8) of \cite{Margalit&Quataert21}. The correct expression is 
$g(\Theta) = \frac{(p-1)\gamma_m}{(p-1)(\gamma_m-1)-p+2} \left(\frac{\gamma_m}{3\Theta}\right)^{p-1}$.
}
The electron temperature is itself a function of the shock velocity, and is determined by Equation~(2) of \cite{Margalit&Quataert21} with $\Theta_0 = \epsilon_T (\mu m_p / \mu_e m_e) (\Gamma-1)$.
In the limit $\Theta \gtrsim 1$ which is relevant for high-velocity shocks, $f(\Theta) \approx g(\Theta) \approx 1$.
The coefficients $C_j(p)$ and $C_\alpha(p)$ are given by Equations~(15,17) of \cite{Margalit&Quataert21}. For $p=3$, these attain the values $C_j = 2/3$ and $C_\alpha \simeq 28.6$. Finally, the function $I^\prime(x)$ is given by Equation~(13) of \cite{Margalit&Quataert21}, and is derived in \cite{Mahadevan+96}.

At the peak of the SED, $\left. dL_\nu/d\nu \right\vert_{\nu_{\rm pk}} = 0$ (by definition). Given observed values of $\nu_{\rm pk}$ and $L_{\nu_{\rm pk}}$, we therefore obtain a set of two equations for two variables, $(\Gamma\beta)_{\rm sh}$, $\dot{M}/v_{\rm w}$. They are
\begin{equation}
\label{eq:Appendix_GeneralEquations}
    L_\nu \left[ \nu_{\rm pk} ~;~ (\Gamma\beta)_{\rm sh},\dot{M}/v_{\rm w} \right] \overset{!}{=} L_{\nu_{\rm pk}}
    ~,~\text{and}~~
    \left(\frac{d L_\nu }{dx}\right) \left[ \nu_{\rm pk} ~;~ (\Gamma\beta)_{\rm sh},\dot{M}/v_{\rm w} \right] \overset{!}{=} 0
.
\end{equation}
In general, these equations must be solved using numerical root-finding procedures.
Note that we have treated the luminosity as a function of the proper-velocity of the shock, $(\Gamma\beta)_{\rm sh}$, whereas several of the expressions above depend instead on the Lorentz factor of post-shock material, $\Gamma$. 
The two are related to one-another via Equation~(5) of \cite{BlandfordMcKee76}. With the effective adiabatic index $\hat{\gamma} = (4+\Gamma^{-1})/3$, this can be rewritten as
\begin{equation}
\label{eq:Appendix_gb_of_gbsh}
    (\Gamma\beta)^2 = \frac{1}{4} \left[ (\Gamma\beta)_{\rm sh}^2 - 2 + \sqrt{ (\Gamma\beta)_{\rm sh}^4 + 5 (\Gamma\beta)_{\rm sh}^2 + 4 } \right] ,
\end{equation}
and recall that $\Gamma = \sqrt{ 1 + (\Gamma\beta)^2 }$.
Note that $(\Gamma\beta)^2 = \frac{9}{16} (\Gamma\beta)_{\rm sh}^2$ in the non-relativistic regime, while $(\Gamma\beta)^2 = \frac{1}{2} (\Gamma\beta)_{\rm sh}^2$ in the ultra-relativistic case. This implies that $(\Gamma\beta)^2 \approx 0.5 (\Gamma\beta)_{\rm sh}^2$ is a reasonable approximation for any velocity.

\section{Critical Luminosity}
\label{sec:Appendix_Lcrit}

In the following we derive more accurate expressions for the limiting peak luminosity, $L_{\rm crit}$.
At the critical luminosity, emission (and absorption) at frequency $\nu_{\rm pk}$ are dominated by thermal electrons. This implies that the terms proportional to $C_j$ or $C_\alpha$ in Equations~(\ref{eq:Appendix_L_general},\ref{eq:Appendix_tau})---which represent the contributions of power-law electrons---can be neglected.
The condition $\left. dL_\nu/d\nu \right\vert_{\nu_{\rm pk}} = 0$ (Equation~\ref{eq:Appendix_GeneralEquations}) then
introduces an equation relating the peak frequency $x_{\rm pk}$ to the optical depth at this frequency, $\tau_{\rm pk} \equiv \tau(x_{\rm pk})$,
which can be written as
\begin{equation}
\label{eq:Appendix_taupk}
    \frac{\tau_{\rm pk}}{e^{\tau_{\rm pk}}-1} 
    = 1 + \left.\frac{1 + \left({d\ln I^\prime}/{d\ln x}\right)}{1 - \left({d\ln I^\prime}/{d\ln x}\right)}\right\vert_{x_{\rm pk}}
    .
\end{equation}
This equation must be solved numerically. Once solved however, one can express the peak luminosity (Equation~\ref{eq:Appendix_L_general}) as a function of a single dimensionless variable, $\tilde{\tau}_{\rm pk} \equiv x_{\rm pk} \tau_\Theta = \tau_{\rm pk}(x_{\rm pk}) x_{\rm pk}^2 / I^\prime(x_{\rm pk})$ (Equation~\ref{eq:Appendix_tau}),
\begin{equation}
\label{eq:Appendix_Lpk_xtau}
    L_{\nu_{\rm pk}}(\tilde{\tau}_{\rm pk}) = \frac{2^{59/8} \pi^{17/8}}{3^{9/16}} m_e c^2 \left( \frac{\epsilon_T f}{\epsilon_B}\right)^{1/8} \left(\nu_{\rm pk}\bar{t}\right)^{17/8}
    \tilde{\tau}_{\rm pk}^{-1/8} \left( 1 - e^{-\tau_{\rm pk}} \right)
    \left(\frac{4-1/\Gamma^2}{8+1/\Gamma^2} \beta \right)^{17/8} \Gamma^3
    .
\end{equation}
Note that in the exponent, $\tau_{\rm pk} = \tau_{\rm pk}(\tilde{\tau}_{\rm pk})$ through Equation~(\ref{eq:Appendix_taupk}).
The Lorentz factor of post-shock material $\Gamma = \Gamma(\tilde{\tau}_{\rm pk})$ is also a function of $\tilde{\tau}_{\rm pk}$ through the following relation
\begin{equation}
\label{eq:Appendix_Gamma_equation}
    \left(\frac{4-1/\Gamma^2}{8+1/\Gamma^2} \sqrt{1-1/\Gamma^2} \right)^{-1/8} \left( \Gamma - 1 \right)  = 
    \frac{\mu_e m_e}{\mu m_p}
    \left(\frac{3^{7/2} 8\pi f \nu_{\rm pk}\bar{t}}{\epsilon_B \epsilon_T^7 \tilde{\tau}_{\rm pk}}\right)^{1/8}
    .
\end{equation}
In both Equations~(\ref{eq:Appendix_Lpk_xtau},\ref{eq:Appendix_Gamma_equation}) we have implicitly assumed that $\Theta \gtrsim 1$.

The maximum peak luminosity can then be calculated from Equation~(\ref{eq:Appendix_Lpk_xtau}) by finding $L_{\rm crit} = \max_{\tilde{\tau}_{\rm pk}} \left(L_{\nu_{\rm pk}} \left[\tilde{\tau}_{\rm pk}\right] \right)$.
When $\Gamma \gg 1$ then $L_{\nu_{\rm pk}} \propto \tilde{\tau}_{\rm pk}^{-1/2} ( 1 - e^{-\tau_{\rm pk}} )$, which peaks at $\tilde{\tau}_{\rm pk} \simeq 0.1712$ (corresponding to $x_{\rm pk} \simeq 1.506$, and $\tau_{\rm pk} \simeq 0.0587$). Alternatively, when $\Gamma-1 \ll 1$, then $L_{\nu_{\rm pk}} \propto \tilde{\tau}_{\rm pk}^{-4/15} ( 1 - e^{-\tau_{\rm pk}} )$. This peaks at $\tilde{\tau}_{\rm pk} \simeq 2.645$ (and a corresponding $x_{\rm pk} \simeq 2.581$, $\tau_{\rm pk} \simeq 0.169$).
With the aid of these numerical coefficients, analytic solutions may be obtained in the non-relativistic and ultra-relativistic regimes. In particular, one can calculate $\Gamma$ at the critical luminosity using Equation~(\ref{eq:Appendix_Gamma_equation}). The shock four-velocity can then be calculated using Equation~(\ref{eq:Appendix_gb_of_gbsh}). Using this method, we find that the proper-velocity of the shock along the critical curve $L_{\nu_{\rm pk}} = L_{\rm crit}$ is
\begin{equation}
\label{eq:Appendix_bG_crit}
    \left(\Gamma\beta\right)_{\rm sh, crit} =
    \begin{dcases}
        \left(\Gamma\beta\right)_{\rm sh, NR} \equiv 
        1.417 \epsilon_{B,-1}^{-1/15} \left(\frac{\epsilon_T}{0.4}\right)^{-7/15} \left(\frac{f}{3/16}\right)^{1/15} \left(\frac{\nu_{\rm pk}\bar{t}}{5\,{\rm GHz} \times 100\,{\rm d}}\right)^{1/15}
        &,~\left(\Gamma\beta\right)_{\rm sh, crit} \ll 1
        \\
        \left(\Gamma\beta\right)_{\rm sh, UR} \equiv 
        1.174 \epsilon_{B,-1}^{-1/8} \left(\frac{\epsilon_T}{0.4}\right)^{-7/8} \left(\frac{f}{3/16}\right)^{1/8} \left(\frac{\nu_{\rm pk}\bar{t}}{5\,{\rm GHz} \times 100\,{\rm d}}\right)^{1/8}
        &,~\left(\Gamma\beta\right)_{\rm sh, crit} \gg 1
    \end{dcases}
\end{equation}
in the non-relativistic $\left(\Gamma\beta\right)_{\rm sh, crit} \ll 1$ and ultra-relativistic $\left(\Gamma\beta\right)_{\rm sh, crit} \gg 1$ limits, respectively. Clearly, neither limit is strictly appropriate for the typical parameters adopted above. This is not surprising since we expect the transition from a SSA to optically-thin peak to occur around $\left(\Gamma\beta\right)_{\rm sh} \sim 1$.
In such cases one must generally use a numerical procedure to maximize $L_{\nu_{\rm pk}}(\tilde{\tau}_{\rm pk})$ (Equation~\ref{eq:Appendix_Lpk_xtau}) and obtain the shock velocity using Equation~(\ref{eq:Appendix_Gamma_equation}; in general, also using a numerical root-solving scheme).
Alternatively, we find that the solution can be well-approximated by the following interpolating function,
\begin{equation}
\label{eq:Appendix_bG_crit_approx}
    \left(\Gamma\beta\right)_{\rm sh, crit} \approx \left[ \left(\Gamma\beta\right)_{\rm sh, NR}^n + \left(\Gamma\beta\right)_{\rm sh, UR}^n \right]^{1/n}
    ~~,~~ n = 2
\end{equation}
with $\left(\Gamma\beta\right)_{\rm sh, NR}$, $\left(\Gamma\beta\right)_{\rm sh, UR}$ defined in Equation~(\ref{eq:Appendix_bG_crit}) above. This approximation has a maximum error of $< 1\%$.
The critical luminosity can similarly be found using the same procedure. In the non-relativistic and ultra-relativistic limits analytic solutions are possible, and we find that
\begin{equation}
    \label{eq:Appendix_Lcrit}
    L_{\rm crit} =
    \begin{dcases}
        L_{\rm crit, NR} \equiv 
        2.73 \times 10^{30}\,{\rm erg\,s}^{-1}\,{\rm Hz}\, \epsilon_{B,-1}^{-4/15} \left(\frac{\epsilon_T}{0.4}\right)^{-13/15} \left(\frac{f}{3/16}\right)^{4/15} \left(\frac{\nu_{\rm pk}\bar{t}}{5\,{\rm GHz} \times 100\,{\rm d}}\right)^{34/15}
        &,~\left(\Gamma\beta\right)_{\rm sh, crit} \ll 1
        \\
        L_{\rm crit, UR} \equiv 
        1.68 \times 10^{30}\,{\rm erg\,s}^{-1}\,{\rm Hz}\, \epsilon_{B,-1}^{-1/2} \left(\frac{\epsilon_T}{0.4}\right)^{-5/2} \left(\frac{f}{3/16}\right)^{1/2} \left(\frac{\nu_{\rm pk}\bar{t}}{5\,{\rm GHz} \times 100\,{\rm d}}\right)^{5/2}
        &,~\left(\Gamma\beta\right)_{\rm sh, crit} \gg 1
    \end{dcases}
    .
\end{equation}
Again, it is typically the case that $\left(\Gamma\beta\right)_{\rm sh, crit} \sim 1$ and neither of these expressions is strictly appropriate. We therefore employ an interpolating function of a similar form to (\ref{eq:Appendix_bG_crit_approx}) and find that 
\begin{equation}
\label{eq:Appendix_Lcrit_approx}
    L_{\rm crit} \approx \left[ L_{\rm crit, NR}^{1/n} + L_{\rm crit, UR}^{1/n} \right]^{n}
    ~~,~~ n = 2
\end{equation}
provides an excellent approximation to the critical luminosity for any combination of parameters (the variables $L_{\rm crit, NR}$, $L_{\rm crit, UR}$ are defined in Equation~\ref{eq:Appendix_Lcrit} above). This interpolating function has a maximum error of $< 5\%$.
Equation~(\ref{eq:Appendix_Lcrit_approx}) should be used instead of the crude approximation given by Equation~(\ref{eq:L_crit}), which is only accurate at the order of magnitude level, and for the range of microphysical parameters and $\nu_{\rm pk} \bar{t}$ considered in Figure~\ref{fig:PhaseSpace_SSA} (in particular, if $\epsilon_T \ll 1$ or  $\epsilon_B \ll 0.1$, Equation~\ref{eq:L_crit} may be inaccurate by orders of magnitude).

\section{Fast Cooling}
\label{sec:Appendix_FastCooling}

Throughout this paper we have made the simplifying assumption that electrons are slow cooling, that is---that the timescale over which electrons loose their energy (e.g., via synchrotron radiative losses) is long compared to other timescales in the problem. This is typically true at $\sim$GHz frequencies, as we show below, but may be incorrect for particularly luminous events and peak frequencies in the sub-mm band or higher. In this Appendix we therefore derive approximate conditions that delineate the regime where electrons cool quickly through synchrotron radiation and where synchrotron losses are negligible and our slow-cooling assumption holds (barring other cooling mechanisms such as inverse-Compton scattering or free-free emission).
Note that we here adopt the terms `slow-cooling' and `fast-cooling' in referring to the cooling timescale of electrons that produce the observed emission at frequency $\nu_{\rm pk}$, as this is the frequency of interest in this paper.

Because the synchrotron power scales as $\propto \gamma^2$, only electrons whose Lorentz factor exceeds 
\begin{equation}
\label{eq:gamma_cool}
    \gamma_{\rm cool} = \frac{ 6\pi m_e c }{ \sigma_T B^2 \Gamma t_e }
\end{equation}
are able to cool appreciably over a timescale $< t_e$. Here $t_e \sim t$ is the time it takes injected electrons to cross the emission region, i.e. $t_e \approx \Delta R / \beta c$ where $\beta$ is the post-shock velocity (here assumed to be constant throughout the post-shock region). Using our definition for $\Delta R$ (Equation~\ref{eq:DeltaR_prime}) we find that
\begin{equation}
\label{eq:te}
    t_e = \frac{4f}{3} \left(\frac{\Gamma_{\rm sh}}{\Gamma}\right)^2 \left(\frac{\beta_{\rm sh}}{\beta}\right) \bar{t}
    \approx
    \begin{cases}
        \frac{1}{3} \bar{t} \left(\frac{f}{3/16}\right) &, \left(\Gamma\beta\right)_{\rm sh} \ll 1
        \\
        \frac{1}{2} \bar{t} \left(\frac{f}{3/16}\right) &, \left(\Gamma\beta\right)_{\rm sh} \gg 1
    \end{cases}
    .
\end{equation}

Using Equations~(\ref{eq:gamma_cool},\ref{eq:te}) we can calculate the corresponding cooling frequency $\nu_{\rm cool}$ above which synchrotron cooling effects will become important. The requirement that electrons dominating emission at $\nu_{\rm pk}$ be slow cooling is equivalent to the condition $\gamma_e(\nu_{\rm pk}) < \gamma_{\rm cool}$, where $\gamma_e(\nu_{\rm pk})$ is the Lorentz factor of electrons that dominate emission at frequency $\nu_{\rm pk}$.
For power-law electrons 
$\gamma_e(\nu) = \sqrt{4\pi m_e c \nu / 3 eB}$. The requirement that electrons be slow cooling at peak therefore translates into the requirement
\begin{equation}
\label{eq:cooling_condition_pl}
    \text{slow cooling:}~
    L_{\nu_{\rm pk}} > 
    1.4 \times 10^{21}\,{\rm erg \,s}^{-1}\,{\rm Hz}^{-1}\, 
    \left(\frac{\epsilon_e/\epsilon_B}{0.1}\right)^{-2}
    \left(\frac{f}{3/16}\right)^{13/3}
    \left(\frac{\nu_{\rm pk}}{5\,{\rm GHz}}\right)^{38/3} \left(\frac{\bar{t}}{100\,{\rm d}}\right)^{19/3}
    ;~L_{\nu_{\rm pk}} < L_{\rm th}
\end{equation}
where we have implicitly assumed $\left(\Gamma\beta\right)_{\rm sh} \ll 1$ above.
When thermal electrons dominate peak emission and $\nu_{\rm pk} \gg \nu_\Theta$ (as appropriate ``deep'' within case b where $\nu_\Theta \ll \nu_{\rm pk} \simeq \nu_{\rm a}$; that is, when $L_{\rm th} < L_{\nu_{\rm pk}} \ll L_{\rm crit}$) then emission at $\nu_{\rm pk}$ is dominated by the high-frequency ``tail'' of comparatively lower-energy electrons 
whose Lorentz factor scales as $\nu^{1/3}$, specifically $\gamma_e(\nu)/\Theta = ( 2\nu / \nu_\Theta )^{1/3}$
\citep{Margalit&Quataert21}. In this case, the condition that peak emission be in the slow-cooling regime translates into the condition
\begin{equation}
\label{eq:cooling_condition_th_SSA}
    \text{slow cooling:}~
    L_{\nu_{\rm pk}} > 
    2 \times 10^{25}\,{\rm erg \,s}^{-1}\,{\rm Hz}^{-1}\, 
    \epsilon_{B,-1}^{20/33} \left(\frac{\epsilon_T}{0.4}\right)^{-43/33}
    \left(\frac{f}{3/16}\right)^{52/33}
    \left(\frac{\nu_{\rm pk}}{5\,{\rm GHz}}\right)^{190/33} \left(\frac{\bar{t}}{100\,{\rm d}}\right)^{118/33}
    ;~L_{\nu_{\rm pk}} > L_{\rm th}
\end{equation}
and again this approximation assumes that $\left(\Gamma\beta\right)_{\rm sh} \ll 1$. At high luminosities, as one approaches $L_{\rm crit}$ this assumption, along with the assumption that $\nu_{\rm pk} \gg \nu_\Theta$ break down and the equation above is not accurate (even at lower velocities, this is only accurate to within a factor of a few because of the approximate relations used in its derivation; see \S\ref{sec:thermal}).

Equation~(\ref{eq:cooling_condition_th_SSA}) is less restrictive than Equation~(\ref{eq:cooling_condition_pl}). That is, some events that would have been in the fast-cooling regime (at frequency $\nu_{\rm pk}$) had they been interpreted using the standard power-law model, may in fact reside well within the slow-cooling regime if interpreted as emission from thermal electrons. This is primarily due to the fact that the Lorentz factors of electrons contributing to emission at frequency $\nu_{\rm pk}$ is lower in the thermal-electron scenario (because $\nu \propto \gamma_e^3$ in this scenario rather than $\nu \propto \gamma_e^2$; see \citealt{Margalit&Quataert21}). Of course, this is only viable for events where $L_{\nu_{\rm pk}} \gtrsim L_{\rm th}$ where the contribution of thermal electrons indeed dominates.

Finally, when peak emission is optically thin and thermal electrons dominate (case c), the emission at $\nu_{\rm pk}$ is dominated by electrons with Lorentz factors $\gamma_e \approx 2 \Theta$.\footnote{
Specifically, the 
emissivity at frequency $\nu_{\rm pk}$ is proportional to $\propto \int z^2 e^{-z} F(x_{\rm pk}/z^2) dz$ where $x_{\rm pk} \equiv \nu_{\rm pk}/\nu_\Theta \simeq 1.268$ and $z \equiv \gamma_e/\Theta$. The integrand of this function peaks at $z \simeq 2.04$, and is therefore dominated by electrons whose Lorentz factor is $\gamma_e \approx 2 \Theta$. 50\% of the total emission at frequency $\nu_{\rm pk}$ is contributed by electrons with Lorentz factors between $z=1.4$ to $z=3$ (90\% is contributed by electrons between $z=0.84$ to $z=5.2$).
} 
Demanding that $\gamma_e < \gamma_{\rm cool}$, and using the results for the 
ultra-relativistic $\left(\Gamma\beta\right)_{\rm sh} \gg 1$ limit where this case is appropriate, we find that the peak is in the slow-cooling regime if
\begin{equation}
    \text{slow cooling:}~
    L_{\nu_{\rm pk}} < 
    1.8 \times 10^{32}\,{\rm erg \,s}^{-1}\,{\rm Hz}^{-1}\, 
    \epsilon_{B,-1}^{-1} \left(\frac{\epsilon_T}{0.4}\right)^{-3} \left(\frac{\nu_{\rm pk}}{5\,{\rm GHz}}\right) \left(\frac{\bar{t}}{100\,{\rm d}}\right)^{2}
    ,~\text{optically-thin}
    .
\end{equation}
Note that in the relativistic optically-thin scenario, the slow-cooling regime applies to lower-luminosity shocks.
In the radio, optically-thin shocks are almost always slow cooling. However this may not necessarily be the case if one considers peak emission at higher frequencies, e.g. in the optical, X-ray, or gamma-ray bands.


%% Include this line if you are using the \addedtext, \replaced, \deleted

\end{document}